\documentclass[%
 aps,
 jmp,%
 amsmath,amssymb,
 reprint,showpacs%
]{revtex4-1}

\usepackage{graphicx}
\usepackage{dcolumn}
\usepackage{bm}
\usepackage{epstopdf}
\usepackage{color}

\usepackage[normalem]{ulem}

\newcommand{\rad}{a}

\begin{document}

\preprint{AIP/123-QED}

\title[A spherical object in an expanding universe]{
Dynamics of a spherical object of uniform density\\in an expanding universe}
\author{Roshina Nandra}
\email{rn288@mrao.cam.ac.uk}
 \affiliation{Astrophysics Group, Cavendish Laboratory, JJ Thomson Avenue, Cambridge CB3 0HE}
 \affiliation{Kavli Institute for Cosmology, c/o Institute of Astronomy, Madingley Road, Cambridge CB3 0HA}
\author{Anthony N. Lasenby}%
 \email{a.n.lasenby@mrao.cam.ac.uk}
\affiliation{Astrophysics Group, Cavendish Laboratory, JJ Thomson Avenue, Cambridge CB3 0HE}%
\affiliation{Kavli Institute for Cosmology, c/o Institute of Astronomy, Madingley Road, Cambridge CB3 0HA}
\author{Michael P. Hobson}
\email{mph@mrao.cam.ac.uk}
\affiliation{Astrophysics Group, Cavendish Laboratory, JJ Thomson Avenue, Cambridge CB3 0HE}%


\begin{abstract}
We present Newtonian and fully general-relativistic solutions for the
evolution of a spherical region of uniform interior density
$\rho_{\rm i}(t)$, embedded in a background of uniform exterior density
$\rho_{\rm e}(t)$. In both regions, the fluid is assumed to support
pressure. In general, the expansion rates of the two regions,
expressed in terms of interior and exterior Hubble parameters $H_{\rm i}(t)$
and $H_{\rm e}(t)$, respectively, are independent. We consider in detail two
special cases: an object with a static boundary, $H_{\rm i}(t)=0$; and an
object whose internal Hubble parameter matches that of the background,
$H_{\rm i}(t)=H_{\rm e}(t)$. In the latter case, we also obtain fully
general-relativistic expressions for the force required to keep a test
particle at rest inside the object, and that required to keep a test
particle on the moving boundary.  We also derive a generalised form of
the Oppenheimer--Volkov equation, valid for general time-dependent
spherically-symmetric systems, which may be of interest in its own
right.
\end{abstract}

\maketitle

%

\section{Introduction\label{sec:intro}}

In a recent paper, we presented metrics for a point mass residing in
each of a spatially-flat, open and closed expanding
universe \cite{NLH1}.  These were derived using a
tetrad-based procedure in general relativity \cite{GGTGA}, and are
essentially a combination of the Schwarzschild metric and the
Friedmann--Robertson--Walker (FRW) metric for a homogeneous and
isotropic universe.  In particular, we used our metrics to study
particle dynamics outside the central object.  As one might expect
intuitively, for radial motion in the Newtonian limit, the force
acting on a test particle was found in all three cases to comprise of
the usual $1/r^2$ inwards component due to the central mass and a
cosmological component proportional to $r$ that is directed outwards
(inwards) when the expansion of the universe is accelerating
(decelerating).

\begin{figure}
\centering
\fbox{\includegraphics[height=2in,width=2.8in]
{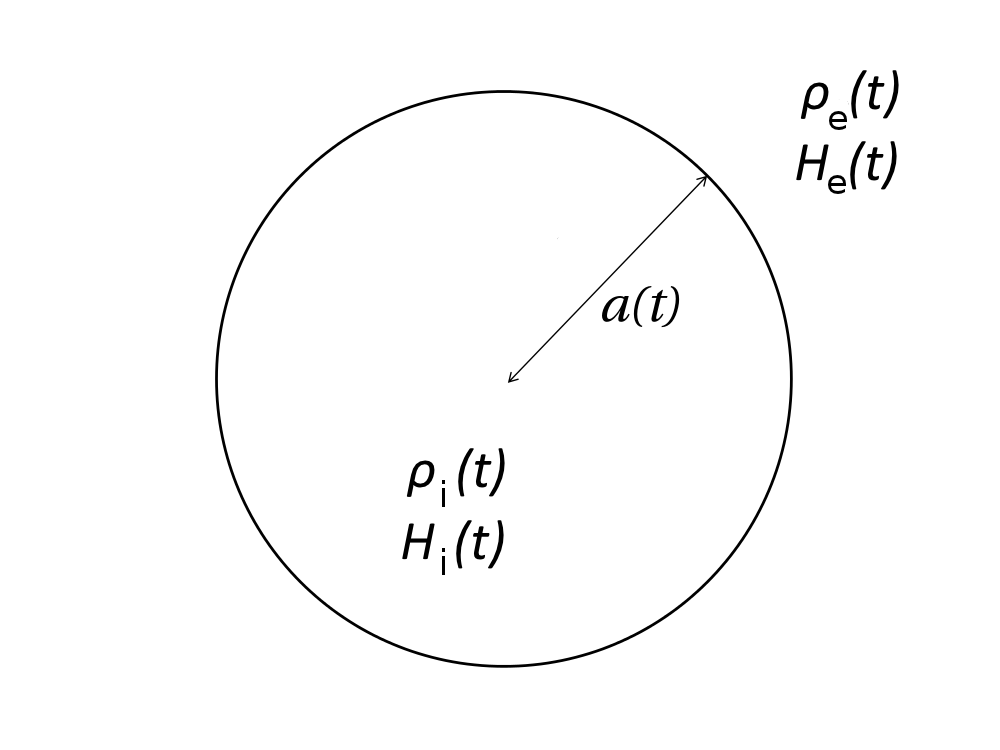}}
  \caption{Model of a spherical object of size $\rad(t)$ and
    uniform interior density $\rho_{\rm i}(t)$ residing in an expanding
    universe with uniform exterior density $\rho_{\rm e}(t)$.  The
    corresponding `Hubble parameters' of the interior and exterior
    regions are denoted $H_{\rm i}(t)$ and $H_{\rm e}(t)$,
    respectively.  \label{fig:setup}}
\end{figure}

A natural progression of this work is to consider a central object of
finite spatial extent. It may initially seem sensible (as it did to
other authors) to proceed by deriving an interior metric that, at the
boundary of the object, matches onto a previously obtained exterior solution
for the case of a central point mass.  In fact, this approach is too
restrictive, since the interior model has to be set up so that
the effects of the object's own expansion vanish at its boundary.  We
therefore consider the problem afresh and model the physical system in
the manner illustrated in Fig.~\ref{fig:setup}. Indeed, this figure
may be taken as the definition of our model, in which a spherical
massive object of size $\rad(t)$ and uniform interior density
$\rho_{\rm i}(t)$ resides in an expanding universe with uniform exterior
density $\rho_{\rm e}(t)$. In general, the spatially-uniform `Hubble
parameters' of the interior and exterior regions are independent, and
denoted by $H_{\rm i}(t)$ and $H_{\rm e}(t)$ respectively.

From the figure, we may write down an expression for the total mass
(or energy in the relativistic case; will use natural units $c=G=1$
throughout) contained within a sphere of physical radius $r$.  We
denote this by $M(r,t)$, but note that dependencies on $r$ and $t$
will usually be suppressed in the equations below, whereas
those on $r$ or $t$ alone will usually be made explicit. It is
clear that
\begin{equation}
  M=\begin{cases}
    \frac{4}{3}\pi\rho_{\rm i}(t)r^3, & \text{$r\leq \rad(t)$},\\
    \frac{4}{3}\pi\rho_{\rm e}(t)(r^3-\rad^3(t))+\frac{4}{3}\pi\rho_{\rm i}(t)\rad^3(t), & \text{$r > \rad(t)$}.\nonumber
  \end{cases}
\end{equation}
We may rewrite the second of these cases and work with the alternative
expressions
\begin{equation}
  M=\begin{cases}
    \frac{4}{3}\pi\rho_{\rm i}(t)r^3, & \text{$r\leq \rad(t)$},\\
   \frac{4}{3}\pi\rho_{\rm e}(t)r^3+m(t), & \text{$r > \rad(t)$},
  \end{cases}	\label{eq:Mdefns}
\end{equation}
where $m(t)=\frac{4}{3}\pi(\rho_{\rm i}(t)-\rho_{\rm e}(t))\rad^3(t)$ is the mass
contained within the boundary $\rad(t)$ at time $t$, {\em in excess} of
that which would be present due to the background alone.

As will we show, in both the Newtonian and fully general-relativistic
cases, the dynamical evolution of the system may be determined
completely by specifying the internal and external Hubble parameters
$H_{\rm i}(t)$ and $H_{\rm e}(t)$, respectively (together with the radius
$a_\ast\equiv a(t_\ast)$ and density $\rho_{\ast}\equiv
\rho_{\rm i}(t_\ast)$ of the object at some reference time $t=t_\ast$).
Typically, we will take $H_{\rm e}(t)$ to correspond to some expanding
exterior universe of interest, but $H_{\rm i}(t)$ can, in principle, have
any form (and be positive or negative). This follows from allowing the
relationship between the fluid pressure and density to be arbitrary,
since then the interplay between the internal pressure of the object
and its self-gravity may allow it to expand or contract at any
rate. This freedom would disappear, however, if one imposed an
equation of state.


We note that, following our approach in \cite{NLH1}, in each region
(interior and exterior) we assume a single `phenomenological' fluid.
This avoids the complexity of an explicit multi-fluid treatment,
whereby one would separate the fluid in each region into its baryonic
and dark matter components.
In particular, in each region we assume a single overall
(uniform) fluid density and a single associated (effective)
pressure. It is envisaged that the pressure comprises the ordinary gas
pressure due to baryonic matter, and an effective pressure from the
dark matter that arises from the motions of dark matter particles
having undergone phase-mixing and relaxation (see \cite{LyndenBell} and
    \cite{binney}).

Our model is sufficiently general to be applied to a range of physical
situations. In reality, no object is simply embedded in the general
fluid of the `expanding universe', but rather inside a hierarchical
collection of objects. By making appropriate choices for the
parameters of the interior and exterior regions, our model could be
used to study a star inside a galaxy, a galaxy inside a cluster or a
cluster inside a supercluster, for example.  It is also possible to
generalise the model to account for objects embedded inside a number
of other objects.  For example, a galaxy embedded inside a cluster,
which itself is embedded in the expanding universe. We will not,
however, pursue this generalisation here.

There have, of course, been numerous previous studies investigating
both the Newtonian and general-relativistic dynamics of
self-gravitating spherical bodies. For example, Misner, Thorne \&
Wheeler \cite{gravitation} describe the spherically-symmetric collapse
of a `ball of dust' having uniform density and zero pressure.  They
later generalise the result to incorporate pressure, but only internal
to the object. A fundamental difference between their work and our
model is that the former considers the exterior spacetime to be static
rather than expanding.  `Swiss cheese' models \cite{harwit} do
incorporate an exterior expanding FRW universe, albeit pressureless,
but the uniform spherical object is surrounded by a `compensating void',
which itself expands into the background and ensures that there is no
net gravitational effect on the exterior universe.  This contrasts
sharply with our model, which does assume the central object to
affect the exterior spacetime.
A more accurate description than the Swiss cheese models is provided
by models based on the Lema\^{i}tre--Tolman--Bondi (LTB) solution
\cite{lemaitre, tolman, bondi}. Such models can incorporate an
arbitrary (usually continuous) density profile for the central object,
which is usually not compensated but can be made so by an appropriate
choice of initial radial density and velocity profiles
\cite{dabrowski1,dabrowski2}. Nonetheless, these models assume both
the interior and exterior regions to be pressureless, although the LTB
solution has recently been extended in \cite{lasky} to describe a
central object with pressure embedded in a static vacuum exterior.


We note that a recent resurgence of interest in Swiss cheese and LTB
models has been prompted by the possibility that they may provide an
explanation for observations of the acceleration of the universal
expansion, without invoking dark energy. This might occur if we, as
observers, reside in a part of the universe that happens to be
expanding faster than the region exterior to it.  By observing a
source in the exterior region, one would then measure an
{\it{apparent}} acceleration of the universe's expansion, but this
would be only a local effect.  The effects of local inhomogeneities on
the apparent acceleration of the universe have been widely studied
\cite{celerier, celerier2, bolejko, bene, kainulainen}, and have been
linked with the observations of distant Type-IA supernova. We
anticipate that our model may also be useful in such a context,
although we leave the investigation of this to future research.

The structure of this paper is as follows.  In Section \ref{sec:newt}
we perform a Newtonian analysis of our model, and derive analytical
expressions for the fluid velocities, densities and pressures in both
the interior and exterior regions. In Section~\ref{sec:framework}, we
then present a full general-relativistic methodology for the analysis
of dynamical spherically-symmetric systems, in which we outline our
tetrad-based approach to solving the Einstein equations, and also
derive a generalised form of the Oppenheimer--Volkov equation, valid
for time-dependent systems. In Section \ref{sec:generalresults}, we
apply this methodology to the analysis of our model system, and obtain
analytical solutions for most of the relevant quantities defining the
line-element in the interior and exterior regions, respectively. For
those quantities that cannot be found analytically, we give the
corresponding differential equations that must be integrated
numerically. As noted above, for a given exterior expanding universe,
the dynamical evolution of the full system is determined by specifying
$H_{\rm i}(t)$. In Section~\ref{sec:static}, we consider the interesting
special case of an object with a static boundary, namely $H_{\rm i}(t)=0$,
and focus particularly on the time-dependent radial pressure profile
in such a system. As our second special case, in
Section~\ref{sec:nolan}, we consider an object for which the internal
Hubble parameter matches that of the background, $H_{\rm i}(t)=H_{\rm e}(t)$. In
this case, we concentrate primarily on the form of the line-element in
the interior and exterior regions, respectively, for spatially-flat,
open and closed background universes. In particular, we show that we
recover the solutions in the exterior region that we obtained
previously in \cite{NLH1}. We also consider the force required to hold
a test particle a rest inside the object, and that required to hold a
test particle on the moving boundary. Finally, we present our
conclusions in Section \ref{sec:conc}.  Throughout this work we
continue to use the subscript $_{\rm i}$ to represent interior quantities
and $_{\rm e}$ to represent exterior quantities.

\section{Newtonian analysis \label{sec:newt}}

Although our ultimate aim is to analyse our model system in
Fig.~\ref{fig:setup} fully general relativistically, it is useful to
begin with a simple Newtonian treatment to develop an intuitive
understanding of its dynamics.

To set up the Newtonian equations, we begin by considering the
velocity potential $\Phi$ and the gravitational potential $V$ for the
interior and exterior regions.  These are related to the fluid
velocity $v$ and gravitational force $F_{\rm g}$ in each region via
the equations $v=\nabla \Phi$ and $F_{\rm g}=-\nabla V$.  The three
equations linking the quantities $\Phi$ and $V$ with the density
$\rho$ and pressure $p$ in each region are the continuity, Euler and
Poisson equations.  Since, in each region,
we have $v=v(r,t)$, $V=V(r,t)$, $p=p(r,t)$ and $\rho=\rho(t)$,
these
equations may be written, with the inclusion of a cosmological
constant $\Lambda$, as
\begin{align}
\frac{\partial \rho(t)}{\partial t}+\frac{\rho(t)}{r^2}\frac{\partial (r^2v)}{\partial r}&=0,&{\text{Continuity}}\nonumber\\
\frac{\partial v}{\partial t}+v\frac{\partial v}{\partial r}+\frac{1}{\rho(t)}\frac{\partial p}{\partial r}+\frac{\partial V}{\partial r}&=0,&{\text{Euler}}\nonumber\\
\frac{1}{r^2}\frac{\partial}{\partial r}\left(r^2\frac{\partial V}{\partial r}\right)&=4\pi \rho(t)-\Lambda.&{\text{Poisson}}\label{eq:Newteqns}
\end{align}

In order to solve this set of equations for the physical quantities
$v$, $\rho$, $p$ and $V$, in both the interior and exterior regions, we
impose four physically reasonable boundary conditions:
\begin{enumerate}
\item both the velocity potential $\Phi$ and gravitational potential
  $V$, and their radial derivatives $v$ and $F_{\rm g}$, match across
  the boundary of the object; \label{item:velbound}
\item the fluid pressure $p$ is continuous across the boundary of the object;
\item all physical quantities behave sensibly throughout the object, so that there are no singularities at the centre, for example; and \label{item:blowup}
\item all physical quantities tend to those of the exterior cosmology
  as $r\rightarrow\infty$.  \label{item:dust}
\end{enumerate}

The three equations in (\ref{eq:Newteqns}) lead to two immediate
observations about the density and pressure across the boundary of the
object.  Although the gradient of $V$ must be continuous across the
boundary, its second-order derivative need not be necessarily.
Therefore, from the Poisson equation, one can see that $\rho$ may
indeed jump at the boundary, as required by our model.  As a
consequence, the Euler equation shows that the gradient of the
pressure may also be discontinuous at the boundary, despite the
pressure itself being continuous there.

We also note that the
continuity of the fluid velocity $v$ across the object boundary, as
specified in our first boundary condition, means that matter does not
cross the boundary in either direction. Thus, the model does not
incorporate accretion onto the object, or outflow away from it.
Hence, the central object's total mass $M(\rad(t),t)$ is constant,
although its excess mass $m(t)$ may vary with time. It is worth
noting, however, that since the pressure is non-zero, work is done as
the boundary moves, so the total internal (thermal) energy of the
object {\em will} change with time.

We may now use the continuity, Euler and Poisson equations to solve
directly for $v$, $\rho$, $V$ and $p$ in both the interior and
exterior regions, in terms of $H_{\rm e}(t)$ and $H_{\rm i}(t)$ (and the object
radius $a_\ast$ and density $\rho_{\ast}$ at some reference time
$t=t_\ast$).

\subsection{General form of the solution}

Since the continuity, Euler and Poisson equations take the form
(\ref{eq:Newteqns}) in both the interior and exterior regions, for the
moment there is no need to distinguish between the regions using the
subscripts $_{\rm i}$ and $_{\rm e}$.

We first obtain an expression for the velocity $v$.  This can be found
directly by integrating the continuity equation, which gives
\begin{equation}
v=-\frac{1}{3}\frac{\rho^{\prime}(t)}{\rho(t)}r+\frac{A_1(t)}{r^2},
\label{eq:v1gensoln}
\end{equation}
where a prime denotes differentiation with respect to the cosmic time
$t$ (we reserve the dot notation for later use in denoting
differentiation with respect to the proper time of observers in our
relativistic treatment), and $A_1(t)$ is an arbitrary function of
$t$. In (\ref{eq:v1gensoln}), we denote the time-dependent factor in
the term proportional to $r$ by $H(t)$, so that
\begin{align}
v&=rH(t)+\frac{A_1(t)}{r^2},\label{eq:vsoln}\\
\rho^{\prime}(t)&=-3\rho(t)H(t).\label{eq:rhoprime}
\end{align}
In each region, the latter equation may be trivially integrated to
obtain an expression for the density $\rho(t)$ in terms of $H(t)$ (and
the density at some reference time $t=t_\ast$).

One may also directly obtain an expression for the gravitational potential $V$
by integrating the Poisson equation, which gives
\begin{equation}
V=\tfrac{1}{6}(4\pi\rho(t)-\Lambda)r^2+\frac{A_2(t)}{r}+A_3(t),\label{eq:V}
\end{equation}
where $A_2(t)$ and $A_3(t)$ are arbitrary functions of time.

Finally, substituting the expressions (\ref{eq:vsoln}) and
(\ref{eq:V}) for $v$ and $V$, respectively, into the Euler equation,
one finds that the solution for $p$ is
\begin{widetext}
\begin{equation}
p=-\left[\tfrac{1}{6}(3H^{\prime}(t)+3H^2(t)+4\pi\rho(t)-\Lambda)r^2
+\frac{A_2(t)+A_1(t)H(t)-A_1'(t)}{r}+\frac{A^2_1(t)}{2r^4}\right]\rho(t)
+A_4(t),\label{eq:ptmp}
\end{equation}
\end{widetext}
where $A_4(t)$  is another arbitrary function of time.

\subsection{Interior region}

Beginning with the interior region, our third boundary condition
requires that there is no singularity at $r=0$, and hence $A_1(t)=0$ in the
expression (\ref{eq:vsoln}) for the fluid velocity. Thus,
\begin{align}
v_{\rm i}&=rH_{\rm i}(t),\label{eq:v1soln}\\
\rho^{\prime}_{\rm i}(t)&=-3\rho_{\rm i}(t)H_{\rm i}(t).\label{eq:v1r1primerho1prime}
\end{align}
Moreover, applying the result (\ref{eq:v1soln}) at the object boundary gives
its rate of growth
\begin{equation}
a'(t) = H_{\rm i}(t)a(t).
\label{eq:boundarygrowth}
\end{equation}
The equations (\ref{eq:v1r1primerho1prime}) and (\ref{eq:boundarygrowth})
may be trivially integrated to obtain the object
radius $a(t)$ and its density $\rho_{\rm i}(t)$ in terms of $H_{\rm i}(t)$ (together with
$a_\ast$ and $\rho_\ast$).

For the gravitational potential $V_{\rm i}$, applying our third boundary
condition again requires that $A_2(t)=0$ in (\ref{eq:V}),
so that one may write
\begin{equation}
V_{\rm i}=V_{\rm b}(t)
-\tfrac{1}{6}(4\pi\rho_{\rm i}(t)-\Lambda)(a^2(t)-r^2),\label{eq:V1}
\end{equation}
where $V_{\rm b}(t)$ is the gravitational potential at the object
boundary, which remains arbitrary.

For the fluid pressure $p_{\rm i}$, applying our third boundary condition once
more requires that $A_1(t)=0=A_2(t)$ in (\ref{eq:ptmp}), so that
\begin{align}
p_{\rm i}&=p_{\rm b}(t)\nonumber \\
&+\tfrac{1}{6}\rho_{\rm i}(t)(3H_{\rm i}^{\prime}(t)+3H_{\rm i}^2(t)+4\pi\rho_{\rm i}(t)-\Lambda)
(a^2(t)-r^2)\label{eq:p1tmp}
\end{align}
where $p_{\rm b}(t)$ is the pressure at the boundary, which can
only be determined after considering the exterior region.

\subsection{Exterior region}

For the exterior region, the arbitrary function $A_1(t)$ in the
expression (\ref{eq:vsoln}) for the fluid velocity need not vanish, so
one may write only
\begin{align}
  v_{\rm e}&=rH_{\rm e}(t)+\frac{A_1(t)}{r^2},\label{eq:v0soln}\\
\rho_{\rm e}^{\prime}(t)&=-3\rho_{\rm e}(t)H_{\rm e}(t).\label{eq:rho0prime}
\end{align}
Nonetheless, we note that (\ref{eq:v0soln}) is consistent with our
fourth boundary condition, which requires that $v_{\rm e}\rightarrow
rH_{\rm e}(t)$ as $r\rightarrow\infty$.  In principle, one could
straightforwardly integrate (\ref{eq:rho0prime}) to obtain an
expression for $\rho_{\rm e}(t)$ in terms of $H_{\rm e}(t)$ and the value of the
external density at some reference time $t=t_\ast$. As we will see,
however, this is unnecessary, since we will shortly obtain an
alternative direct expression for $\rho_{\rm e}(t)$, which is given in
(\ref{eq:rho0H0newt}) below.

For the gravitational potential $v_{\rm e}$, in the exterior region one
cannot require the arbitrary functions $A_2(t)$ and $A_3(t)$ in
(\ref{eq:V}) to vanish, and hence the expression remains
\begin{equation}
v_{\rm e}=\tfrac{1}{6}(4\pi\rho_{\rm e}(t)-\Lambda)r^2+\frac{A_2(t)}{r}+A_3(t).
\label{eq:V0}
\end{equation}

For the fluid pressure $p_{\rm e}$, by applying our fourth boundary
condition, we require that, as $r\to \infty$, the right-hand side of
(\ref{eq:ptmp}) tends to some uniform time-dependent pressure
$p_\infty(t)$ corresponding to the external cosmological model. This
implies that $A_4(t) = p_\infty(t)$ and that the $r^2$ term on the
right-hand side must vanish. The latter condition leads immediately to
the following expression for $\rho_{\rm e}(t)$ in terms of $H_{\rm e}(t)$:
\begin{equation}
H_{\rm e}^{\prime}(t)+H_{\rm e}^2(t)-\tfrac{1}{3}\Lambda= -\tfrac{4}{3}\pi \rho_{\rm e}(t).\label{eq:rho0H0newt}
\end{equation}
This is easily recognised as the standard (general relativistic)
dynamical cosmological field equation, written in terms of the
external Hubble parameter, with the exception that the right-hand side
does not depend on the fluid pressure in addition to the density. This
is to be expected, however, since, as is well known, Newtonian theory
does not take into account the {\em gravitational} effect of the fluid
pressure. The resulting expression for the fluid pressure is then
\begin{equation}
p_{\rm e} = p_\infty(t)-\left[\frac{A_2(t)+A_1(t)H_{\rm e}(t)-A_1'(t)}{r}+\frac{A_1^2(t)}{2r^4}\right]
\rho_{\rm e}(t).\label{eq:p0tmp}
\end{equation}

\subsection{Matching at the object boundary}

To complete our analysis, all that remains is to match the interior
and exterior solutions at the object boundary according to our chosen
conditions there.

Demanding that the expressions (\ref{eq:v1soln}) and (\ref{eq:v0soln})
for the fluid velocities $v_{\rm i}$ and $v_{\rm e}$, respectively, match at the
boundary immediately gives an expression for $A_1(t)$, such that the
exterior fluid velocity becomes
\begin{equation}
v_{\rm e}=rH_{\rm e}(t)-\frac{\rad^3(t)}{r^2}(H_{\rm e}(t)-H_{\rm i}(t)).\label{eq:v0newt}
\end{equation}
This elegant expression clearly shows how the standard cosmological
result is modified by the presence of the central object, unless
$H_{\rm i}(t)=H_{\rm e}(t)$. This modification was not present in the
point mass analysis in \cite{NLH1}, indicating the unrealistic nature
of that model.

Applying the conditions that the gravitational potential and its
radial derivative must match across the boundary yields expressions for
$A_2(t)$ and $A_3(t)$ in (\ref{eq:V0}), which allows us to rewrite the exterior gravitational potential as
\begin{align}
v_{\rm e}&=V_{\rm b}(t)
+\tfrac{1}{6}(4\pi\rho_{\rm e}(t)-\Lambda)(r^2-a^2(t))\nonumber \\
&\hspace{1.25cm}+\tfrac{4}{3}\pi \rad^2(t)\left(1-\frac{a(t)}{r}\right)
(\rho_{\rm i}(t)-\rho_{\rm e}(t)),\label{eq:V0final}
\end{align}
where $V_{\rm b}(t)$ remains undetermined.

Finally, one may insert the expressions obtained for $A_1(t)$ and
$A_2(t)$ into the expression (\ref{eq:p0tmp}) for the exterior
pressure $p_{\rm e}$. Moreover, demanding that the pressure is continuous
across the boundary allows one also to derive a form for $p_{\rm
  b}(t)$. Combining these results, keeping $\rho_{\rm e}(t)$ and $\rho_{\rm i}(t)$
in our expressions for brevity, and momentarily dropping the
$t$-dependencies, the expressions (\ref{eq:p1tmp}) and (\ref{eq:p0tmp})
for the interior and exterior
pressure, respectively, may be written as
\begin{widetext}
\begin{align}
p_{\rm i}&=p_\infty+\rho_{\rm e} a^2\left[\tfrac{4}{3}\pi(\rho_{\rm i}-\rho_{\rm e})+(H_{\rm i}-H_{\rm e})' + (H^2_{\rm i}-H^2_{\rm e})+\tfrac{3}{2}(H_{\rm i}-H_{\rm e})^2\right] +\tfrac{1}{6}\rho_{\rm i}\left(4\pi\rho_{\rm i}
+3H^{\prime}_{\rm i}+3H_{\rm i}^2-\Lambda\right)(a^2-r^2),\label{eq:Newtonianpressureint}\\[0.1cm]
p_{\rm e}&=p_\infty
+\rho_{\rm e}\rad^3
\left[\tfrac{4}{3}\pi(\rho_{\rm i}-\rho_{\rm e})+(H_{\rm i}-H_{\rm e})'+(H^2_{\rm i}-H^2_{\rm e})+2(H_{\rm i}-H_{\rm e})^2\right] r^{-1}-\tfrac{1}{2}\rho_{\rm e}\rad^6\left(H_{\rm e}-H_{\rm i}\right)^2r^{-4}.\label{eq:Newtonianpressureext}
\end{align}
\end{widetext}

We now have a complete Newtonian solution for our model system, which
can be written in terms of the internal and external Hubble parameters
$H_{\rm i}(t)$ and $H_{\rm e}(t)$, respectively (and $a_\ast$ and $\rho_\ast$). The resulting
expressions are expected to be reasonably accurate for a wide range of
physical systems that resemble our model.  Nonetheless, for an exact
analysis it is necessary to employ the equations of general
relativity.

\section{General relativistic methodology}
\label{sec:framework}

We now describe our general-relativistic methodology for the analysis
of spherically-symmetric systems. We solve the Einstein field
equations for such systems using the tetrad-based method described in
\cite{NLH1}, and originally presented in \cite{GGTGA}, which we now
summarise.

\subsection{Tetrad-based solution for spherical systems}

In a Riemannian spacetime in which events are labelled with a set of
coordinates $x^{\mu}$, each point has the corresponding coordinate
basis vectors $\mathbf{e}_{\mu}$, related to the metric via
$\mathbf{e}_{\mu}\cdot\mathbf{e}_{\nu}=g_{\mu\nu}$.  At each point we
may also define a {\itshape{local Lorentz frame}} by another set of
orthogonal basis vectors $\hat{\mathbf{e}}_i$ (Roman indices), which
are not derived from any coordinate system and are related to the
Minkowski metric $\eta_{ij}=\mbox{diag}(1,-1,-1,-1)$ via
$\hat{\mathbf{e}}_i\cdot \hat{\mathbf{e}}_j=\eta_{ij}$.  One can
describe a vector $\mathbf{v}$ at any point in terms of its components
in either basis: for example $v_{\mu}=\mathbf{v}\cdot \mathbf{e}_{\mu}$
and $\hat{v}_i=\mathbf{v}\cdot\hat{\mathbf{e}}_{i}$.  The relationship
between the two sets of basis vectors is defined in terms of tetrads,
or {{vierbeins}} ${e_{k}}^{\mu}$, where the inverse is denoted
${e^{k}}_{\mu}$:
\begin{align}
\hat{\mathbf{e}}_k&={e_{k}}^{\mu}\mathbf{e}_{\mu},\nonumber\\
\mathbf{e}_{\mu}&={e^{k}}_{\mu}\hat{\mathbf{e}}_k.
\end{align}
It is not difficult to show that the metric elements are given in
terms of the tetrads by $g_{\mu\nu}=\eta_{ij}e^i_{\ \mu}e^j_{\ \nu}$.

The local Lorentz frames at each point define a family of ideal
observers whose worldlines are the integral curves of the timelike
unit vector field $\hat{\mathbf{e}}_0$. Along a given worldline, the
three spacelike unit vector fields $\hat{\mathbf{e}}_i$ $(i=1,2,3)$
specify the spatial triad carried by the corresponding observer. The
triad may be thought of as defining the orthogonal spatial coordinate
axes of a local laboratory frame that is valid very near the
observer's worldline. In general, the worldlines need not be time-like
geodesics, and hence the observers may be accelerating.

For our spherically-symmetric system, we work in terms of the tetrad
components $f_1\equiv {e_0}^0$, $g_1\equiv {e_1}^1$ and $g_2\equiv
{e_0}^1$, as described in \cite{NLH1}, which define the system via the
line-element
\begin{equation}
ds^2=\frac{g_1^{2}-g_2^{2}}{f_1^{2}g_1^{2}}\,dt^2+\frac{2g_2}{f_1g_1^{2}}\,dt\,dr-\frac{1}{g_1^{2}}\,dr^2-r^2d\Omega^2,\label{eq:generalmetric}
\end{equation}
where $d\Omega$ is an element of solid angle and we have adopted a
`physical' (non-comoving) radial coordinate $r$, whereby the proper
surface area of a sphere of radius $r$ is given by $4\pi r^2$.  We
have also made use of the invariance of general relativity under
general coordinate transformations to choose a time coordinate $t$
that specialises to the so-called Newtonian gauge ($f_2\equiv
{e_1}^0=0$; see \cite{GGTGA}).  It is also convenient to introduce
explicitly the spin-connection coefficients $F\equiv {\omega^0}_{11}$
and $G\equiv {\omega^1}_{00}$, as described in \cite{NLH1}; since we
are assuming standard general relativity, for which torsion vanishes,
the spin-connection can, however, be written entirely in terms of the
tetrad components and their derivatives.  Each quantity is, in
general, a function of $t$ and $r$.

For matter in the form of a perfect fluid with proper density $\rho$
and isotropic rest-frame pressure $p$, the equations linking the
quantities $f_1$, $g_1$, $g_2$, $F$ and $G$ have been shown to be \cite{GGTGA}
\begin{align}
L_rf_1&=-Gf_1\Rightarrow f_1=\exp\left\{-\int^r\frac{G}{g_1}dr\right\},\nonumber\\
L_r g_1&=F g_2+\frac{M}{r^2}-\tfrac{1}{3}\Lambda r-4\pi r
\rho,\nonumber\\
L_r p&=-G(\rho+p),\nonumber\\
L_rM&=4\pi g_1 r^2\rho,\nonumber\\
L_t g_2&=G g_1-\frac{M}{r^2}+\tfrac{1}{3}\Lambda r-4\pi r p,\nonumber\\
L_t\rho&=-\left(\frac{2g_2}{r}+F\right)(\rho+p),\nonumber\\
L_t M&=-4\pi g_2 r^2p,\label{eq:alleqns}
\end{align}
where the two linear differential operators are defined by
\begin{align}
L_t&\equiv f_1\partial_t+g_2\partial_r,\nonumber\\
L_r&\equiv g_1\partial_r,\label{eq:operators}
\end{align}
and the function $F$, radial acceleration $G$ and mass $M$ (or energy)
contained within some radius $r$ are defined via
\begin{align}
L_t g_1&\equiv G g_2,\nonumber\\
L_r g_2 &\equiv F g_1,\nonumber\\
M&\equiv \tfrac{1}{2}r\left(g_2^{2}-g_1^{2}+1-\tfrac{1}{3}\Lambda r^2\right),\label{eq:quantities}
\end{align}
where $\Lambda$ is the cosmological constant.  The above equations
provide a means of uniquely determining the quantities describing a
spherically symmetric physical system, given a specific form chosen
for $M$.  In the next section we will solve the equations using our
definition for $M$ describing a finite, uniform-density object
embedded in an expanding universe, given by equation
(\ref{eq:Mdefns}).

We note that in deriving the system of equations (\ref{eq:alleqns}),
we have made use of the invariance of general relativity under local
rotations of the Lorentz frames to choose the timelike unit vector
$\hat{\mathbf{e}}_0$ at each point to coincide with the four-velocity
of the fluid at that point. Thus, by construction, the four-velocity
$\mathbf{u}$ of an observer comoving with the fluid (or a fluid
particle) has components $[\hat{u}^i] = [1,0,0,0]$ in the tetrad
frame. Since $u^{\mu}=e_i^{\ \mu}\hat{u}^i$, the four-velocity may be
written in terms of the tetrad components and the coordinate basis
vectors as $\mathbf{u}= f_1 \mathbf{e}_0 + g_2 \mathbf{e}_1$. Thus,
the components of a comoving observer's four-velocity in the
coordinate basis are simply $[u^\mu] \equiv
[\dot{t},\dot{r},\dot{\theta},\dot{\phi}]=[f_1,g_2,0,0]$, where dots
denote differentiation with respect to the observer's proper time
$\tau$.  Consequently, we may identify the differential operator $L_t$
in (\ref{eq:operators}) as the derivative with respect to the proper
time of a comoving observer, since $L_t = \dot{t}\partial_t +
\dot{r}\partial_r = d/d\tau$ (it is also straightforward to show that
$L_r$ coincides with the derivative with respect to the radial proper
distance of a comoving observer).

Moreover, since $g_2$ is the rate of
change of the $r$ coordinate of a comoving observer (or fluid
particle) with respect to its proper time, it can be physically
interpreted as the fluid velocity, which we denoted by $v$ in the
Newtonian analysis. We will therefore, in general, use $g_2$ and $v$
interchangeably in our general-relativistic analysis. We also note
that the $L_t\rho$ equation may thus be regarded as the general
relativistic equivalent of the continuity equation given in
(\ref{eq:Newteqns}).

Finally, as shown in \cite{NLH1}, the proper
radial acceleration of a comoving observer (or fluid particle) is $G$,
and hence the motion is, in general, not geodesic.  This behaviour
results from the presence of a pressure gradient in the fluid; indeed
the $L_rp$ equation in (\ref{eq:alleqns}) shows that, in the absence
of a pressure gradient, $G$ vanishes and hence the motion becomes
geodesic.

\subsection{Densities, pressures and forces}

It is worth noting that, as shown in \cite{NLH1}, one can derive general
expressions in terms of the five functions $f_1$, $g_1$, $g_2$, $F$ and $G$
for important physical quantities.  These expressions
can be applied to any system for which $M$ is specified.

Assuming a matter energy-momentum tensor describing a perfect
fluid, the following expressions give the density and pressure of the
fluid in terms of the quantities defining the metric
(\ref{eq:generalmetric}):
\begin{align}
8\pi\rho&=\frac{1}{r^2}(g_2^2-g_1^2 + 1 - \Lambda r^2)
+ \frac{1}{r}\partial_r(g^2_2 - g^2_1),\nonumber\\
8\pi p&=-\frac{f_1}{rg_2}\partial_t(g^2_2-g^2_1) - 8\pi \rho.\label{eq:generalpandrho}
\end{align}
The form for $p$ appears to be analytical, but since $f_1$ is defined
through an integral, for any system of interest a suitable boundary
condition would be required to fix the form for $p$ completely.  It
may therefore sometimes be more instructive simply to leave the
expression for the pressure in terms of the differential equation
defining it.

In \cite{NLH1} we also derived an expression for the force required to
keep a test particle at rest relative to the origin; the idealised
particle was assumed to be infinitesimal, and so not subject to fluid
forces due to pressure gradients, but only to
gravitational forces.  As discussed in \cite{NLH1}, the relationship
between the `physical' coordinate $r$ and proper radial distance
$\ell$ to origin is defined through $d\ell = -(1/g_1)\,dr$, so it is
only in cases for which $g_1 = g_1(r)$ is independent of $t$ that
$\dot{r}=0$ or $\dot{\ell}=0$ are equivalent conditions (where the dot
denotes differentiation with respect to the particle's proper
time). In general, one must thus choose which condition defines `at
rest'. As in \cite{NLH1}, here we adopt the condition $\dot{r}=0$,
which corresponds physically to keeping the test particle on the
surface of a sphere with proper area $4\pi r^2$.  In practice, it
would probably be easier for an astronaut (i.e. test particle) to make
a local measurement to determine the proper area of the sphere on
which he is located, rather than to determine his proper distance to
the origin. With this proviso, the required force, which is the
negative of the force experienced by such a particle, is given by
\begin{equation}
f =
\frac{1}{\sqrt{g_1^{2}-g_2^{2}}}\left[
\frac{f_1g_1(g_2\partial_tg_1-g_1\partial_t g_2)}{g_1^{2}-g_2^{2}}+
  Gg_1-Fg_2\right].\label{eq:finalforcegen2}
\end{equation}

\subsection{Generalised Oppenheimer--Volkov equation}\label{sec:OV}

It is possible to combine some of the equations in (\ref{eq:alleqns})
to obtain a differential equation for the radial pressure gradient
$\partial_r p$ in terms of the pressure $p$, the fluid
velocity $v$ ($\equiv g_2$) and $M$ (or, equivalently, $\rho$ via the
$L_rM$ equation).  In the special case where all the quantities are
independent of $t$, and hence functions of $r$ alone, the resulting
equation should reduce to the standard Oppenheimer--Volkov equation
for a static spherically-symmetric system \cite{oppenheimer}.
Accounting for the possible time-dependence of the quantities,
however, makes our result valid for any spherically-symmetric system,
so we refer to it as a `generalised Oppenheimer--Volkov' equation.  We
will use this equation later to obtain forms for the interior and
exterior pressures for our model system, but our general equation may
also be of interest in its own right.

From the $L_rp$ equation in (\ref{eq:alleqns}), one has
\begin{equation}
\partial_r p =-\frac{G}{g_1}(\rho+p).\label{eq:OVgen}
\end{equation}
For this to be considered a generalised Oppenheimer--Volkov equation,
we require forms for both $g_1$ and $G$.  One may immediately write
down an expression for the former in terms of $v$ and $M$ using the
definition of $M$ in equation (\ref{eq:quantities}):
\begin{equation}
g_1=\sqrt{1-\frac{2M}{r}+v^2-\tfrac{1}{3}{\Lambda}r^2}.\label{eq:g1general}
\end{equation}
Obtaining an expression for $G$ in terms of $v$ and $M$ alone is
slightly more complicated.  Combining the $L_t g_1$ equation in
(\ref{eq:quantities}) with equation (\ref{eq:g1general}) and the $L_rM$ equation in
(\ref{eq:alleqns}), gives
\begin{equation}
G=\frac{\frac{f_1}{v}\left(v\partial_t v-\frac{1}{r}\partial_t M\right)
+\frac{M}{r^2}-4\pi  r \rho +v\partial_r v-\tfrac{1}{3}{\Lambda}r}
{\sqrt{1-\frac{2M}{r}+v^2-\tfrac{1}{3}\Lambda r^2}}.
\label{eq:Gsofar}
\end{equation}
One can then eliminate $f_1$ from this expression, effectively replacing it
with $p$, using the $L_tM$ equation in (\ref{eq:alleqns}), which
implies a form for $f_1$ given by
\begin{equation}
f_1=-4\pi r^2v(\rho+p)\left(\partial_t M \right)^{-1}.\label{eq:f1_1}
\end{equation}
The final generalised Oppenheimer--Volkov equation is obtained by substituting equations (\ref{eq:g1general}), (\ref{eq:Gsofar}) and (\ref{eq:f1_1}) into (\ref{eq:OVgen}) to give
\begin{widetext}
\begin{equation}
\partial_r p = - \left(\frac{\rho+p}{r}\right)\frac{M+4\pi r^3p-\frac{1}{3}\Lambda r^3 + r^2 v\partial_r v - 4\pi r^4 (\rho+p)  (\partial_t M)^{-1}v\partial_t v}
{(1+v^2)r-2M-\frac{1}{3}\Lambda r^3}.\label{eq:OVgeneralised}
\end{equation}
\end{widetext}

We illustrate the use of this result in Appendix A by obtaining the
relationship between $p$ and $\rho$ for the particular case of a point
mass in a homogeneous and isotropic expanding universe, as studied in
\cite{NLH1}. We also note that in the special case of a stationary object,
$v=0$ and the $t$-dependency in (\ref{eq:OVgeneralised}) is lost; it
can then easily be seen that our result reduces, as required, to the
standard Oppenheimer--Volkov equation with a cosmological constant
\cite{winter}:
\begin{equation}
\frac{dp(r)}{dr}=-\left[\frac{\rho(r)+p(r)}{r}\right]
\frac{M(r)+4\pi r^3p(r)-\tfrac{1}{3}{\Lambda}r^3}{r-2M(r)-\tfrac{1}{3}{\Lambda}r^3}.
\label{eq:staticOV}
\end{equation}

\section{General relativistic analysis}\label{sec:generalresults}

We now apply the general-relativistic methodology outlined above to
the analysis of our model system depicted in Fig.~\ref{fig:setup}, for
which $M$ is specified by (\ref{eq:Mdefns}).  This entails solving the
corresponding equations (\ref{eq:alleqns}) and (\ref{eq:quantities}),
in the interior and exterior regions, for the tetrad components
$f_1$, $g_1$ and $g_2$, thus obtaining a form for the spacetime metric
(\ref{eq:generalmetric}), and the spin-connection coefficients $F$ and
$G$ (which can be written in terms of the tetrad components and their
derivatives). From now on, we will exclusively use $v$ instead of
$g_2$, and distinguish between interior and exterior quantities using
the subscripts $_{\rm i}$ and $_{\rm e}$ respectively, so that the interior
quantities are written as $f_{1,{\rm i}}$, $g_{1,{\rm i}}$, $v_{\rm i}$, $F_{\rm i}$ and
$G_{\rm i}$ and the corresponding exterior quantities are written as
$f_{1,{\rm e}}$, $g_{1,{\rm e}}$, $v_{\rm e}$, $F_{\rm e}$ and $G_{\rm e}$.

\subsection{Boundary conditions}

In the general-relativistic case, on the 3-dimensional (timelike)
hypersurface $\Sigma$ traced out by the (in general) moving spherical
boundary of the object, the solution must satisfy the Israel junction
conditions \citep{Israel66,Israel67}. These require continuity both of
the induced metric and the extrinsic curvature on $\Sigma$. As shown
in Appendix B, the first condition requires continuity of $f_1$, $g_1$
and $v$, and the second condition further requires the continuity of
$\partial_r f_1$.

Using the equations (\ref{eq:alleqns}), the above junction conditions
have consequences for the continuity of other variables of
interest. In particular, from the $L_rf_1$ equation, one has $G =
-(g_1/f_1)\partial_r f_1$, which must therefore also be continuous at
the object boundary. Moreover, the $L_rp$ equation and the continuity
of $g_1$ and $G$ imply that the pressure $p$ is also continuous across
the boundary, although its radial derivative, in general, has a step
there.

Comparing the relativistic boundary conditions with those adopted in
the Newtonian analysis in Section~\ref{sec:newt}, the continuity of $v$ and
the radial acceleration $G$ are analogous to the continuity of $v$ and
the Newtonian force $F_{\rm g}$ assumed in boundary condition
1. Moreover, the continuity in $p$ recovers boundary condition 2. In
our general relativistic analysis below, we will also again adopt
condition 3 that there are no singularities in physical quantities.

We note that, as in the Newtonian case, the continuity of the fluid
velocity $v$ at the object boundary means that matter does not cross
the boundary in either direction, and so the model does not
incorporate accretion onto the object, or outflow away from it.  In
the general-relativistic case, however, this does {\em not} imply that
$M(a(t), t)$ is constant, since this now denotes the object's total
energy, which does change as the boundary moves, as is clear from the
$L_tM$ and $L_rM$ equations in (\ref{eq:alleqns}).

Finally, we also adopt the general relativistic equivalent of our
previous boundary condition 4, namely that all physical quantities
tend to those of the exterior cosmology for large $r$. For
spatially-flat and open universes, this corresponds to $r\to \infty$,
whereas for a closed universe one must instead consider the limit
$a(t) \ll r < R(t)$, where $R(t)$ is the universal scale factor which
also corresponds to the curvature scale for a closed universe. In each
case, we require the line-element (\ref{eq:generalmetric}) to tend to
the corresponding FRW line-element at large $r$. Our `physical' radial
coordinate $r$ is simply related to the usual `comoving' radial
coordinate $\hat{r}$ by $r=\hat{r}R(t)$, so that the FRW line-element
becomes
\begin{align}
ds^2 &= \frac{[1-\frac{kr^2}{R^2(t)}-r^2H_{\rm e}^2(t)]\,dt^2
+ 2rH_{\rm e}(t)\,dt\,dr - dr^2}
{1-\frac{kr^2}{R^2(t)}}\nonumber \\
&\hspace{5.5cm}-r^2\,d\Omega^2,
\end{align}
where $H_{\rm e}(t)$ is the exterior Hubble parameter and $k=0,-1,1$
corresponds to a spatially-flat, open or closed exterior cosmology,
respectively. Comparing this line-element with that in
(\ref{eq:generalmetric}), one quickly sees that, for large $r$, one
requires
\begin{align}
f_1 &\to 1,\nonumber \\
g_1 &\to \sqrt{1-\frac{kr^2}{R^2(t)}},\nonumber\\
g_2 &\to rH_{\rm e}(t).\label{eq:bclarger}
\end{align}

\subsection{Interior region}\label{sec:g1vFgeneral}

Let us begin by considering the interior region, for which
$M=\frac{4}{3}\pi\rho_{\rm i}(t)r^3$. As in the Newtonian case, we begin by
finding an expression for $v_{\rm i}$.  The $L_tM$ and $L_t\rho$ equations
in (\ref{eq:alleqns}) may be combined with the above expression for
$M$ and the $L_rg_2$ equation from (\ref{eq:quantities}) to obtain
\begin{equation}
\partial_r v_{\rm i} =\frac{v_{\rm i}}{r},\nonumber
\end{equation}
which is immediately solved for $v_{\rm i}$ to yield
\begin{align}
v_{\rm i}&=rH_{\rm i}(t),\label{eq:g2int}
\end{align}
where $H_{\rm i}(t)$ is the `constant' of integration which we have defined
as the interior Hubble parameter. Hence, one also has the simple
relation $F_{\rm i} \equiv \partial_r v_{\rm i} = H_{\rm i}(t)$.

Substituting the above expressions for $v_{\rm i}$ and $F_{\rm i}$ back into the
$L_t\rho$ equation in (\ref{eq:alleqns}), gives the useful result
\begin{equation}
L_t\rho_{\rm i}(t)=-3H_{\rm i}(t)(\rho_{\rm i}(t)+p_{\rm i}).
\label{eq:Ltrho1}
\end{equation}
This is analogous to the Newtonian expression in
(\ref{eq:v1r1primerho1prime}), but with $\rho_{\rm i}(t)$ on the right-hand side replaced by $\rho_{\rm i}(t)+p_{\rm i}$, and $\rho_{\rm i}'(t)$ on the left-hand
side replaced by $L_t\rho_{\rm i}(t)$; as discussed previously, $L_t$
corresponds simply to the derivative with respect to the proper time
of an observer moving with the fluid. Using (\ref{eq:Ltrho1}), we may write
\begin{equation}
f_{1,{\rm i}} = -\frac{3H_{\rm i}(t)(\rho_{\rm i}(t)+p_{\rm i})}{\rho_{\rm i}'(t)}.
\label{eq:f1intsol}
\end{equation}

Moreover, substituting our expressions for $v_{\rm i}$ and $M$ into
(\ref{eq:quantities}), immediately gives
\begin{equation}
g^2_{1,{\rm i}}=1+\left(H_{\rm i}^2(t)-\tfrac{8}{3}\pi\rho_{\rm i}(t)-\tfrac{1}{3}\Lambda
\right)r^2.\label{eq:g1int}
\end{equation}

In order to evaluate the above expressions for $f_{1,{\rm i}}$ and
$g_{1,{\rm i}}$, one requires forms for $\rho_{\rm i}(t)$ and $p_{\rm i}$.  Unlike in
the Newtonian case, however, one cannot obtain analytical solutions
for these functions. Nonetheless, substituting the above expressions
for $M$ and $v_{\rm i}$ into the generalised Oppenheimer--Volkov equation
(\ref{eq:OVgeneralised}), and integrating, will yield an (integral)
expression for $p_{\rm i}$ in terms of $H_{\rm i}(t)$, $\rho_{\rm i}(t)$ and the pressure on
the boundary $p_{\rm b}(t)$, where the last two functions can only be
determined after considering the exterior region.
This should be
contrasted with the Newtonian case, for which only the determination
of $p_{\rm b}(t)$ first required consideration of the exterior region.

\subsection{Exterior region}

For the exterior region, one has $M=\frac{4}{3}\pi\rho_{\rm e}(t)r^3 +
m(t)$. Indeed, since both $\rho_{\rm e}(t)$ and $\rho_{\rm i}(t)$ appear in the
exterior definition of $M$, it is initially ambiguous to which density
the symbol `$\rho$' in equations (\ref{eq:alleqns}) refers.
Using the $L_rM$ equation, however, one quickly finds that
`$\rho\text{'}=\rho_{\rm e}(t)$.

We again begin by finding an expression for the velocity.  Using the
same procedure as in the interior region, one finds that
\begin{equation}
\partial_r v_{\rm e} =\frac{2v_{\rm e}}{r}\left(\frac{2\pi r^3\rho_{\rm e}^{\prime}(t)-3m^{\prime}(t)}{4\pi r^3\rho_{\rm e}^{\prime}(t)+3m^{\prime}(t)}\right),\nonumber
\end{equation}
which may easily be integrated to obtain the solution
\begin{equation}
v_{\rm e}=\frac{A(t)}{r^2}(4\pi r^3\rho_{\rm e}^{\prime}(t)+3m^{\prime}(t)),\nonumber
\end{equation}
where $A(t)$ is an arbitrary function of time. This function may be determined
using the boundary condition $v_{\rm e}\rightarrow rH_{\rm e}(t)$ for large $r$
from (\ref{eq:bclarger}), which gives
\begin{equation}
v_{\rm e}=rH_{\rm e}(t)+\frac{3m^{\prime}(t)H_{\rm e}(t)}{4\pi r^2\rho_{\rm e}^{\prime}(t)},
\label{eq:g2ext}
\end{equation}
from which the corresponding expression for $F_{\rm e} =\partial_r v_{\rm e}$ is easily
obtained.

Substituting these expressions for $v_{\rm e}$ and $F_{\rm e}$ back into the
$L_t\rho$ equation in (\ref{eq:alleqns}), one obtains the useful result
$L_t\rho_{\rm e}(t)=-3H_{\rm e}(t)(\rho_{\rm e}(t)+p_{\rm e})$. This is analogous to that
found in the interior region and immediately leads to
\begin{equation}
f_{1,{\rm e}} = -\frac{3H_{\rm e}(t)(\rho_{\rm e}(t)+p_{\rm e})}{\rho_{\rm e}'(t)}.
\label{eq:f1extsol}
\end{equation}

Moreover, substituting the expressions for $M$ and $v_{\rm e}$ into
(\ref{eq:quantities}) then gives
\begin{align}
g^2_{1,{\rm e}} &=
1-\frac{2m(t)}{r}
+\left(H_{\rm e}^2(t)-\tfrac{8}{3}\pi\rho_{\rm e}(t)-\tfrac{1}{3}\Lambda\right)r^2
\nonumber \\
&\hspace{0.65cm}+\frac{9H_{\rm e}^2(t)m'(t)}{16\pi^2\rho_{\rm e}^{\prime 2}(t) r^4}
\left(\tfrac{8}{3}\pi \rho_{\rm e}'(t)r^3 + m'(t)\right).
\label{eq:g2extgensol}
\end{align}

The boundary conditions (\ref{eq:bclarger}) on $f_1$ and $g_1$ at
large $r$ therefore require
\begin{align}
\rho_{\rm e}'(t) + 3H_{\rm e}(t)(\rho_{\rm e}(t)+p_\infty(t)) &=
0, \label{eq:cosmo1}\\ H_{\rm e}^2(t)-\tfrac{8}{3}\pi\rho_{\rm e}(t)-\tfrac{1}{3}\Lambda
&= -\frac{k}{R^2(t)}\label{eq:cosmo2},
\end{align}
where the uniform time-dependent pressure $p_\infty(t)$ corresponds to
the external cosmological model. If desired, one may define the
equation-of-state parameter $w_\infty(t)\equiv p_\infty(t)/\rho_{\rm e}(t)$,
which would typically be time-independent for standard cosmological
fluids such as dust ($w_\infty=0$) or radiation ($w_\infty=1/3$). We
recognise (\ref{eq:cosmo1}) and (\ref{eq:cosmo2}) as the standard
cosmological fluid evolution equation and the Friedmann equation,
respectively. Moreover, these can be combined in the usual manner to
yield the dynamical cosmological field equation
\begin{equation}
H'_{\rm e}(t) + H_{\rm e}^2(t) -\tfrac{1}{3}\Lambda
= -\tfrac{4}{3}\pi(\rho_{\rm e}(t)+3p_\infty(t)),
\label{eq:cosmo3}
\end{equation}
which thus provides an expression for $\rho_{\rm e}(t)$ in terms of $H_{\rm e}(t)$
(and the assumed form for $p_\infty(t)$ or $w_\infty(t)$).

In order to evaluate the above expressions for $f_{1,{\rm e}}$ and
$g_{1,{\rm e}}$, one also requires forms for $p_{\rm e}$, $\rho_{\rm
  i}(t)$ and $a(t)$. It is not possible, in general, to obtain
analytical solutions for these quantities. Nonetheless, substituting
our expressions for $M$ and $v_{\rm e}$ into the generalised
Oppenheimer--Volkov equation (\ref{eq:OVgeneralised}), integrating and
using (\ref{eq:cosmo3}), will yield an (integral) expression for
$p_{\rm e}$ in terms of $\rho_{\rm i}(t)$, $a(t)$, $p_{\infty}(t)$ and
$H_{\rm e}(t)$, which may then be used to obtain the boundary pressure
$p_{\rm b}(t)=p_{\rm e}(a(t),t)$ in terms of these functions.  One is
free to specify $p_{\infty}(t)$ and $H_{\rm e}(t)$, and the functions
$\rho_{\rm i}(t)$ and $a(t)$ may be determined by considering the
matching of the interior and exterior solutions at the object
boundary.

\subsection{Matching at the object boundary}

To complete the solution in both the interior and exterior regions one
requires $\rho_{\rm i}(t)$ and $a(t)$. This may be achieved by
imposing the required junction conditions at the object boundary. As
discussed earlier, we may satisfy the junction conditions by requiring
$f_1$, $g_1$, $v$ and $p$ to be continuous there; indeed, the last
condition has already been assumed in defining the boundary pressure
$p_{\rm b}(t)$.

If $f_1$ and $p$ are continuous at the boundary, the
expressions (\ref{eq:f1intsol}) and (\ref{eq:f1extsol}) immediately
yield the relationship
\begin{equation}
\rho_{\rm i}^{\prime}(t)=\frac{H_{\rm i}(t)\rho_{\rm e}^{\prime}(t)}{H_{\rm e}(t)}
\left(\frac{\rho_{\rm i}(t)+p_{\rm b}(t)}{\rho_{\rm e}(t)+p_{\rm b}(t)}\right).
\label{eq:rho1prime}
\end{equation}
The above expression
should be contrasted with the corresponding Newtonian result
(\ref{eq:v1r1primerho1prime}). In the latter, $\rho_{\rm i}(t)$ is
determined entirely by $H_{\rm i}(t)$ (and
$\rho_\ast$), whereas in the general-relativistic case $\rho_{\rm i}(t)$
also depends on $p_{\rm b}(t)$, and on
the exterior Hubble parameter $H_{\rm e}(t)$ and density $\rho_{\rm e}(t)$. This
shows that the exterior spacetime has an effect on the dynamics of the
interior, contrary to popular belief. As recently discussed
by \cite{zhang}, this opinion stems from a common misunderstanding of
Birkhoff's theorem.

If $v$ is continuous at the boundary, one first notes from the general
expression (\ref{eq:g1general}) that $g_1$ is also automatically
continuous there. Moreover, combining (\ref{eq:g2int}) and
(\ref{eq:g2ext}), the continuity of $v$ yields an expression for the
rate of growth of the boundary, given by
\begin{equation}
\rad^{\prime}(t)=-\frac{\rad(t)H_{\rm i}(t)\rho_{\rm e}^{\prime}(t)}{3H_{\rm e}(t)(\rho_{\rm e}(t)+p_b(t))}.
\label{eq:r1prime}
\end{equation}
We note that in deriving
this expression we have also made use of the result
(\ref{eq:rho1prime}) and that $m(t)=\frac{4}{3}
\pi(\rho_{\rm i}(t)-\rho_{\rm e}(t)) a^3(t)$. By evaluating the expression
(\ref{eq:f1extsol}) at the object boundary, one sees that
(\ref{eq:r1prime}) may be written as $f_{1,b}(t)\,a'(t)=H_{\rm i}(t)a(t)$,
which is equivalent to the elegant result
\begin{equation}
L_t a(t) = H_{\rm i}(t) a(t),
\label{eq:lta}
\end{equation}
where the operator $L_t$ is, in this case, evaluated at the object
boundary. Comparing this expression with the Newtonian result
(\ref{eq:boundarygrowth}) for the rate of growth of the boundary, one
sees that $a'(t)$ on the left-hand side of the latter has simply been
replaced by $L_ta(t)$, which corresponds to the derivative with
respect to the proper time of an observer comoving with the fluid.
This observation also reconciles (\ref{eq:lta}) with the expression
(\ref{eq:g2int}) for the fluid velocity in the interior
region. Evaluating the latter at the boundary and recalling that $v
(\equiv g_2)$ is the rate of change of the $r$ coordinate of a fluid
particle with respect to its proper time, one sees that it is indeed
consistent with (\ref{eq:lta}). This also verifies our earlier
conclusion that the central object experiences no accretion
or outflow. It is also worth noting that, in the Newtonian case,
$a(t)$ depends only on $H_{\rm i}(t)$ (and
$a_\ast$), whereas the general-relativistic expression
(\ref{eq:r1prime}) also depends on the boundary pressure $p_{\rm b}(t)$, and on
the exterior Hubble parameter $H_{\rm e}(t)$ and density
$\rho_{\rm e}(t)$.  This provides another example of the exterior spacetime
influencing the dynamics of the interior.

We note that substituting our expression for $m(t)$ into
(\ref{eq:g2ext}), and using the results (\ref{eq:rho1prime}) and
(\ref{eq:r1prime}), leads to the elegant expression
\begin{equation}
v_{\rm e}=rH_{\rm e}(t)-\frac{\rad^3(t)}{r^2}(H_{\rm e}(t)-H_{\rm i}(t)),
\label{eq:v0elegant}
\end{equation}
which is identical to that found in (\ref{eq:v0newt}) from the
Newtonian analysis; thus, there is no general-relativistic correction
to the fluid velocity. In a similar manner, one finds that the
expression (\ref{eq:g2extgensol}) becomes (momentarily suppressing
$t$-dependencies for brevity)
\begin{align}
g^2_{1,{\rm e}} &=
1-\frac{2m}{r}
+\left(H_{\rm e}^2-\tfrac{8}{3}\pi\rho_{\rm e}-\tfrac{1}{3}\Lambda\right)r^2
\nonumber \\
&\hspace{0.65cm}+\frac{a^3}{r}
(H_{\rm i}-H_{\rm e})
\left[2H_{\rm e}+\frac{a^3}{r^3}(H_{\rm i}-H_{\rm e})\right].
\label{eq:g1elegant}
\end{align}

To complete the solution, the differential equations
(\ref{eq:rho1prime}) and (\ref{eq:r1prime}) must typically be
(numerically) integrated simultaneously to obtain $\rho_{\rm i}(t)$
and $a(t)$, since, in general, the expression for $p_{\rm b}(t)$
depends on both these functions.

\medskip
Finally, we have arrived at a full solution to our system of equations
(albeit not analytical), with all quantities specified in terms of the
interior and exterior Hubble parameters $H_{\rm i}(t)$ and $H_{\rm
  e}(t)$ (together with the object density $\rho_\ast$ and radius
$a_\ast$ at some reference time $t=t_\ast$, and the pressure
$p_\infty(t)$ or equation-of-state parameter $w_\infty(t)$ of the
external cosmological fluid at large $r$). For a given exterior
cosmology and initial conditions, the solution thus depends only on
the internal Hubble parameter $H_{\rm i}(t)$, which we are free to
choose.  To illustrate our solution, we now consider two special
cases: $H_{\rm i}(t)=0$, which corresponds to an object with a static
boundary; and $H_{\rm i}(t)=H_{\rm e}(t)$, which corresponds to an
object whose expansion rate coincides with that of the background.

\section{Object with $H_{\rm i}(t)=0$}
\label{sec:static}

For an object with $H_{\rm i}(t)=0$, one sees immediately from
(\ref{eq:r1prime}) that the boundary is static, namely
$\rad^{\prime}(t)=0$, or equivalently $\rad(t)=\rad=\mbox{constant}$.
Intuitively, one would expect an object with a fixed boundary to have
a constant total energy.  This is confirmed by equation
(\ref{eq:rho1prime}), which implies that
$\rho_{\rm i}(t)=\rho_{\rm i}=\mbox{constant}$, so that the total energy of the
object from equation (\ref{eq:Mdefns}) is $M(a(t),t)=(4/3)\pi\rho_{\rm i}
\rad^3=\mbox{constant}$, which we will denote by ${\cal M}$, but
its excess mass (energy) $m(t)$ is still time-dependent.  At first
glance, this physical situation appears to mirror the Schwarzschild
interior uniform-density solution.  Differences arise, however,
because of the presence of the expanding fluid exterior to the object
in our model, rather than a simple vacuum as in the Schwarzschild
case.  The expanding exterior fluid causes the boundary conditions at
the edge of the object to be time-dependent, rather than fixed, and
therefore changes the overall interior and exterior solutions.

\subsection{Interior region}

In the interior region, $M = \tfrac{4}{3}\pi\rho_{\rm i} r^3$. From
(\ref{eq:g2int}), one confirms immediately that $v_{\rm i}=0$, so the entire
fluid in the interior of the object is static. As a result, the
line-element (\ref{eq:generalmetric}) in the interior region reduces
to the simple diagonal form
\begin{equation}
ds^2=\frac{1}{f_{1,{\rm i}}^{2}}dt^2-\frac{1}{g_{1,{\rm i}}^{2}}\,dr^2-r^2d\Omega^2.
\label{eq:metric1}
\end{equation}

From (\ref{eq:g1int}), one immediately finds
\begin{equation}
g_{1,{\rm i}}(r) = \sqrt{1-(\tfrac{8}{3}\pi\rho_{\rm i}+\tfrac{1}{3}\Lambda)r^2},
\label{eq:eg1-g1sol}
\end{equation}
which we note is time-independent. In order to obtain a form for
$f_{1,{\rm i}}$, it is convenient first to find an expression for $p_{\rm i}$, using
the generalised Oppenheimer--Volkov equation
(\ref{eq:OVgeneralised}). Since $v_{\rm i}=0$, this reduces to the form
(\ref{eq:staticOV}), but with $dp/dr$ on the left-hand side replaced
by $\partial_r p$, since the interior pressure still depends on
$t$, and reads
\begin{equation}
\partial_r p_{\rm i}=-\frac{(\rho_{\rm i}+p_{\rm i}) (\tfrac{4}{3}\pi\rho_{\rm i}+4\pi p_{\rm i} -\tfrac{1}{3}\Lambda)r}
{1-(\tfrac{8}{3}\pi\rho_{\rm i} +\tfrac{1}{3}\Lambda)r^2}.
\label{eq:p1prime}
\end{equation}
Recalling that $p_{\rm i}=p_{\rm i}(r,t)$ and $\rho_{\rm i}$ is a constant, this
equation is separable and can be integrated straightforwardly to
obtain
\begin{align}
&\frac{4\pi\rho_{\rm i}+12\pi p_{\rm i}-\Lambda}{\rho_{\rm i}+p_{\rm i}}\nonumber\\
&\indent=\frac{4\pi\rho_{\rm i}+12\pi p_c(t)-\Lambda}{\rho_{\rm i}+p_c(t)}\sqrt{1
-(\tfrac{8}{3}\pi\rho_{\rm i}+\tfrac{1}{3}\Lambda)r^2},\label{eq:intpressurestatic}
\end{align}
where $p_{\rm c}(t)$ denotes the time-dependent central pressure of
the object. We note that $p_{\rm c}(t)$ could be eliminated, and
effectively replaced with the boundary pressure $p_{\rm b}(t)$, by
remembering that $p_{\rm i}(\rad(t),t)=p_{\rm b}(t)$.  We choose not to do
this here, however, since we will later use the expression in its
current form.  It is also worth noting that, if desired, equation
(\ref{eq:intpressurestatic}) can be reorganised into an explicit
expression for $p_{\rm i}$, but the result is somewhat complicated and
unilluminating, so we retain the implicit form
(\ref{eq:intpressurestatic}).

Since $H_{\rm i}(t)=0$ and $\rho_{\rm i}'(t)=0$, our general form
(\ref{eq:f1intsol}) for $f_{1,{\rm i}}$ is undefined, so we must use an
alternative expression in this case. Combining the $L_r f_1$ and $L_r
p$ equations in (\ref{eq:alleqns}), one quickly finds the general result
\begin{equation}
\partial_r \ln f_1 = \frac{\partial_r p}{\rho+p}.
\end{equation}
Thus, integrating this equation directly and using (\ref{eq:p1prime}),
in this case one has
\begin{equation}
f_{1,{\rm i}} = A(t)
\exp\left[-\int^r \frac{(\tfrac{4}{3}\pi\rho_{\rm i}+4\pi p_{\rm i} -\tfrac{1}{3}\Lambda)
\bar{r}}
{1-(\tfrac{8}{3}\pi\rho_{\rm i} +\tfrac{1}{3}\Lambda)\bar{r}^2}\,d\bar{r}\right],
\end{equation}
where $A(t)$ is an arbitrary function of time. This integral may be performed
analytically using (\ref{eq:intpressurestatic}) to obtain
\begin{widetext}
\begin{equation}
f_{1,{\rm i}} = f_{1,b}(t)
\frac
{4\pi(\rho_{\rm i}+p_{\rm c}(t))-(\tfrac{4}{3}\pi\rho_{\rm i}+4\pi p_{\rm c}(t)-\tfrac{1}{3}\Lambda)
\sqrt{1-(\tfrac{8}{3}\pi\rho_{\rm i}+\tfrac{1}{3}\Lambda)a^2}}
{4\pi(\rho_{\rm i}+p_{\rm c}(t))-(\tfrac{4}{3}\pi\rho_{\rm i}+4\pi p_{\rm c}(t)-\tfrac{1}{3}\Lambda)
\sqrt{1-(\tfrac{8}{3}\pi\rho_{\rm i}+\tfrac{1}{3}\Lambda)r^2}},
\label{eq:eg1-f1sol}
\end{equation}
\end{widetext}
where we have written the solution explicitly in terms of the boundary
value $f_{1,b}(t)$ and the central pressure $p_{\rm c}(t)$ (which can
itself be expressed in terms of the pressure $p_{\rm b}(t)$ at the
boundary, if desired). These functions can be determined only after
considering the exterior region. Note that the expressions
(\ref{eq:eg1-g1sol}) and (\ref{eq:eg1-f1sol}) may be trivially
rewritten in terms of the constant mass ${\cal
  M}=\tfrac{4}{3}\pi\rho_{\rm i}a^3$ of the object, rather than its density
$\rho_{\rm i}$.

\subsection{Exterior region}

In the exterior region, $M={\cal
  M}+\tfrac{4}{3}\pi\rho_{\rm e}(t)(r^3-a^3)$.  The expression
(\ref{eq:v0elegant}) for the fluid velocity $v_{\rm e}$ becomes
\begin{equation}
v_{\rm e}=rH_{\rm e}(t)\left(1-\frac{\rad^3}{r^3}\right),\label{eq:v0static}
\end{equation}
and, from (\ref{eq:g1elegant}), one quickly finds
\begin{align}
g^2_{1,{\rm e}} &= 1-\frac{2{\cal M}}{r}-\tfrac{1}{3}\Lambda r^2 \nonumber \\
&+ \left[H_{\rm e}^2(t)\left(1-\frac{a^3}{r^3}\right)-\tfrac{8}{3}\pi\rho_{\rm e}(t)\right]
\left(1-\frac{a^3}{r^3}\right)r^2.
\end{align}

To obtain a form for $f_{1,{\rm e}}$, one must again first consider the
generalised Oppenheimer--Volkov equation (\ref{eq:OVgeneralised}) to
find an expression for $p_{\rm e}$. Substituting our expressions for $M$ and
$v_{\rm e}$ into (\ref{eq:OVgeneralised}) leads to an equation for
$\partial_r p_{\rm e}$ in terms of $p_{\rm e}$, $\rho_{\rm i}$ and $a$, together with
$\rho_{\rm e}(t)$ and $H_{\rm e}(t)$ and their time-derivatives. For our later
development, however, it is useful to eliminate $\rho_{\rm e}^{\prime}(t)$,
$H_{\rm e}(t)$ and $H_{\rm e}^{\prime}(t)$ in favour of $\rho_{\rm e}(t)$ using the
exterior cosmological equations
(\ref{eq:cosmo1})--(\ref{eq:cosmo3}). This can be performed for a
general exterior cosmology, but for the sake of simplicity we will
restrict our attention to a spatially-flat model $(k=0)$ in which the
cosmological fluid is dust $(p_\infty=0)$, which provides a reasonable
approximation to our current Universe. In this case, one obtains
\begin{widetext}
\begin{equation}
\partial_r p_{\rm e} =
-\left(\frac{\rho_{\rm e}(t)+p_{\rm e}}{r}\right)\frac{\tfrac{4}{3}\pi a^3
\left[\rho_{\rm i}+4\rho_{\rm e}(t)\left(1-\frac{a^3}{r^3}\right)\right]
+4\pi a^3 p_{\rm e} +\tfrac{1}{3}\Lambda a^3\left(1-\frac{2a^3}{r^3}\right)}
{r-\tfrac{8}{3}\pi a^3\left[\rho_{\rm i}+\rho_{\rm e}(t)\left(1-\frac{a^3}{r^3}\right)\right]
-\tfrac{1}{3}\Lambda a^3\left(2-\frac{a^3}{r^3}\right)},
\label{eq:eg1OV}
\end{equation}
\end{widetext}
which, in general, must be integrated numerically to obtain $p_{\rm e}$. One
may then find $f_{1,{\rm e}}$ using (\ref{eq:f1extsol}).

By writing the Oppeheimer--Volkov equation in the form
(\ref{eq:eg1OV}), however, we may obtain approximate analytical
solutions for the special case of a dense object, for which $\rho_{\rm i}\gg
\rho_{\rm e}(t)$. First, we consider the case of a vanishing cosmological
constant $\Lambda = 0$, so that the exterior cosmology is
Einstein--de-Sitter (EdS). Remembering that $r\geq \rad$ in the
exterior, in this case (\ref{eq:eg1OV}) becomes simply
\begin{equation}
\partial_r p_{\rm e} \approx
-\left(\frac{\rho_{\rm e}(t)+p_{\rm e}}{r}\right)
\frac{{\cal M}+4\pi a^3 p_{\rm e}}{r(1-2{\cal M}/r)},
\label{eq:OVapprox}
\end{equation}
which is separable and thus straightforward to solve for
$p_{\rm e}$. Moreover, if one again applies the condition $\rho_{\rm i}\gg
\rho_{\rm e}(t)$ in the resulting solution, one finds that the exterior pressure can be written in
the concise approximate form
\begin{equation}
p_{\rm e}\approx \rho_{\rm e}(t)\left(\frac{1}{\sqrt{1-\frac{2{\cal M}}{r}}}-1\right).
\label{eq:approxpext}
\end{equation}
Second, we consider the case of a more general spatially-flat exterior
cosmology with $\Lambda \neq 0$, but in the limit that $a \to 0$ and
$\rho_{\rm i} \to \infty$ in such as way that the object mass ${\cal M}$
remains constant. In this case, the Oppenheimer--Volkov equation
(\ref{eq:eg1OV}) reduces to (\ref{eq:OVapprox}) but without the $4\pi
a^3 p_{\rm e}$ term on the right-hand side, to which the solution is again
given by (\ref{eq:approxpext}). In each case, the pressure $p_{\rm b}(t)$ at
the boundary $r=a$ can be read off from this equation, which in turn allows
the central pressure $p_{\rm c}(t)$ of the object to be determined
from (\ref{eq:intpressurestatic}).

Moreover, in each case, substituting (\ref{eq:approxpext}) into the
expression (\ref{eq:f1extsol}) for $f_{1,{\rm e}}$, and using the relation
(\ref{eq:cosmo1}) with $p_\infty=0$, one immediately finds the
time-independent result
\begin{equation}
f_{1,{\rm e}} \approx \frac{1}{\sqrt{1-\frac{2{\cal M}}{r}}},
\end{equation}
from which one can read off the value $f_{\rm 1,b}$ at the boundary,
which in turn allows one to specify $f_{1,{\rm i}}$ using (\ref{eq:eg1-f1sol}).

Perhaps unsurprisingly, the above results for $p_{\rm e}$ and $f_{1,{\rm e}}$ are
identical to those external to a point mass ${\cal M}$ embdedded in a
spatially-flat universe, as derived in \cite{NLH1}; however,
${\cal M}$ now represents the mass of a finite-size object.

\subsection{Application to the Sun}

We illustrate these results in two concrete situations. An obvious
first example, though not one where we would expect to see any effects
of significance, is to a star embedded in an external expanding
universe. We take as our example the Sun, and model the pressure
interior and exterior to the Sun, under the (crude) assumptions that
its interior density is uniform and that it is embedded simply in an
expanding homogeneous universe.

The Sun has ${\cal M} \approx 2\times10^{30}$~kg and $\rad\approx
7\times 10^8$~m and, for the purpose of illustration, we will assume a
spatially-flat concordance $\Lambda$CDM exterior cosmology with
$p_{\infty}=0$. In this case, the exterior Hubble parameter is given
by \citep{GRbook}
\begin{equation}
H_{\rm e}(t) = \sqrt{\tfrac{1}{3}\Lambda}
\coth\left(\tfrac{3}{2}\sqrt{\tfrac{1}{3}\Lambda}\,t\right),\label{eq:LambdaCDM}
\end{equation}
and we adopt the observed value of the cosmological constant
$\Lambda \approx 10^{-35}$~s$^{-2}$ \citep{WMAP}. The corresponding expression
for $\rho_{\rm e}(t)$ can be inferred from (\ref{eq:cosmo2}), and is
easily shown to be
\begin{equation}
\rho_{\rm e}(t) = \frac{\Lambda}{8\pi}\,\mbox{cosech}^2
\left(\tfrac{3}{2}\sqrt{\tfrac{1}{3}\Lambda}\,t\right).
\end{equation}

At the current epoch, the external density and pressure are very much
smaller than the internal density and central pressure in the
Sun. Additionally, the Sun is far from being in a solution regime
where general-relativistic corrections are significant for the
pressure profile. Thus both exact and approximate general-relativistic
methods, as well as the Newtonian solutions
(\ref{eq:Newtonianpressureint}) and (\ref{eq:Newtonianpressureext})
found earlier, will all give indistinguishable results for the
interior and exterior pressure profiles. These results are still
interesting, however. The key feature is that the external pressure
lifts the whole internal pressure curve up uniformly, by an amount
equal to the pressure at the boundary. We can see this from the exact
result (\ref{eq:intpressurestatic}), where if we rewrite this in terms
of the pressure at the boundary, $p_{\rm b}(t)$, instead of at the
centre, $p_{\rm c}(t)$, and then carry out an expansion in which
$\rho_{\rm i}$ is treated as a first order quantity, and $p_{\rm
  b}(t)$ and $\Lambda$ are treated as second order quantities, we get
the following result, accurate at second order:
\begin{equation}
p_{\rm i} \approx 
\frac{3 {\cal M}^2}{8\pi a^4}\left(1-\frac{r^2}{a^2}\right)+p_{\rm b}(t).
\label{eq:approx-uplift}
\end{equation}
This is the ordinary Newtonian pressure profile for a uniform-density
object, lifted up by the excess pressure at the boundary. This
boundary pressure can be calculated from (\ref{eq:approxpext})
evaluated at $r=a$, and changes with time due to the expansion of the
universe.

\begin{figure*}
\centering
\includegraphics[height=2in,width=3in]{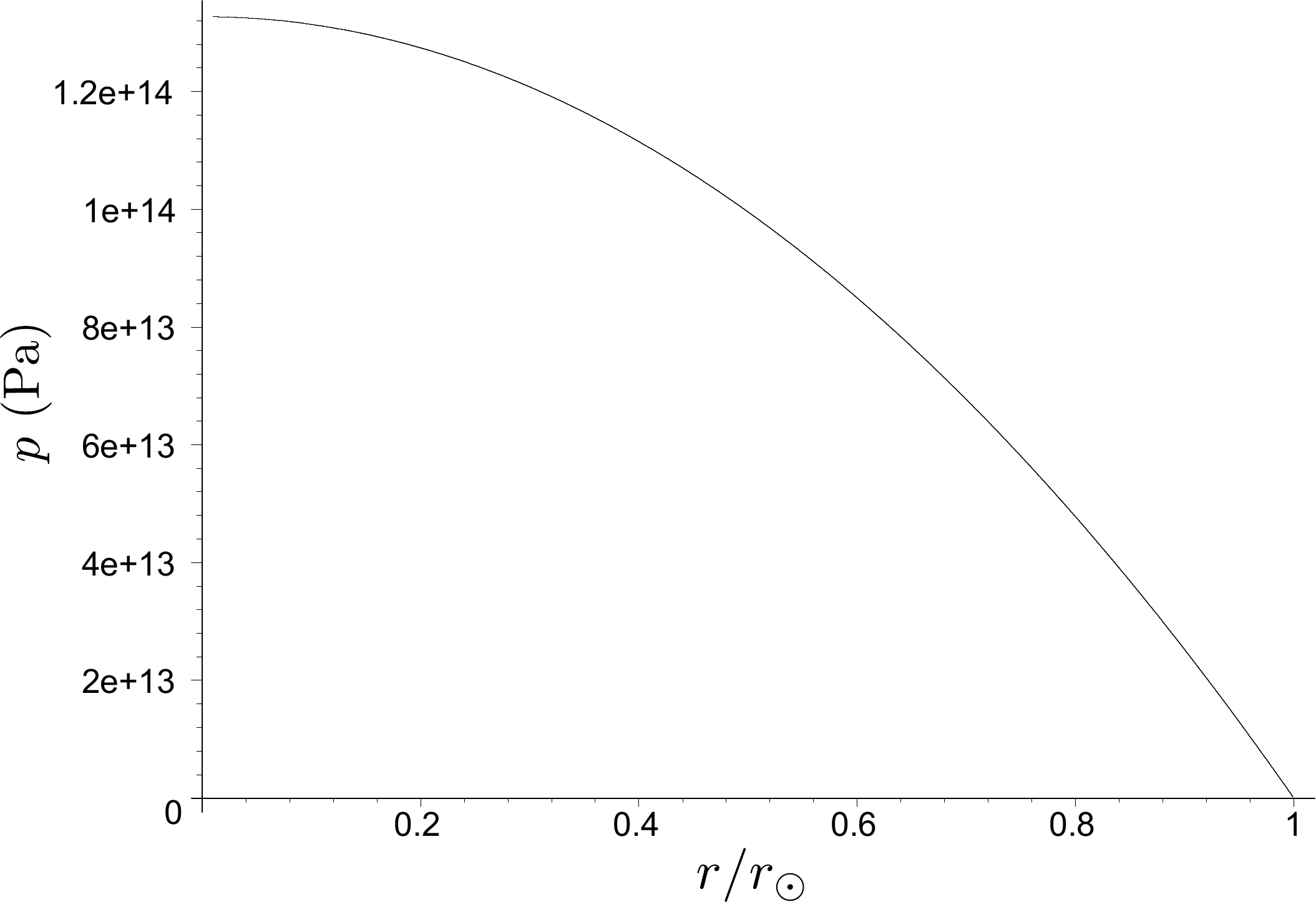}
\hspace{1.5cm}
\includegraphics[height=2in,width=3in]{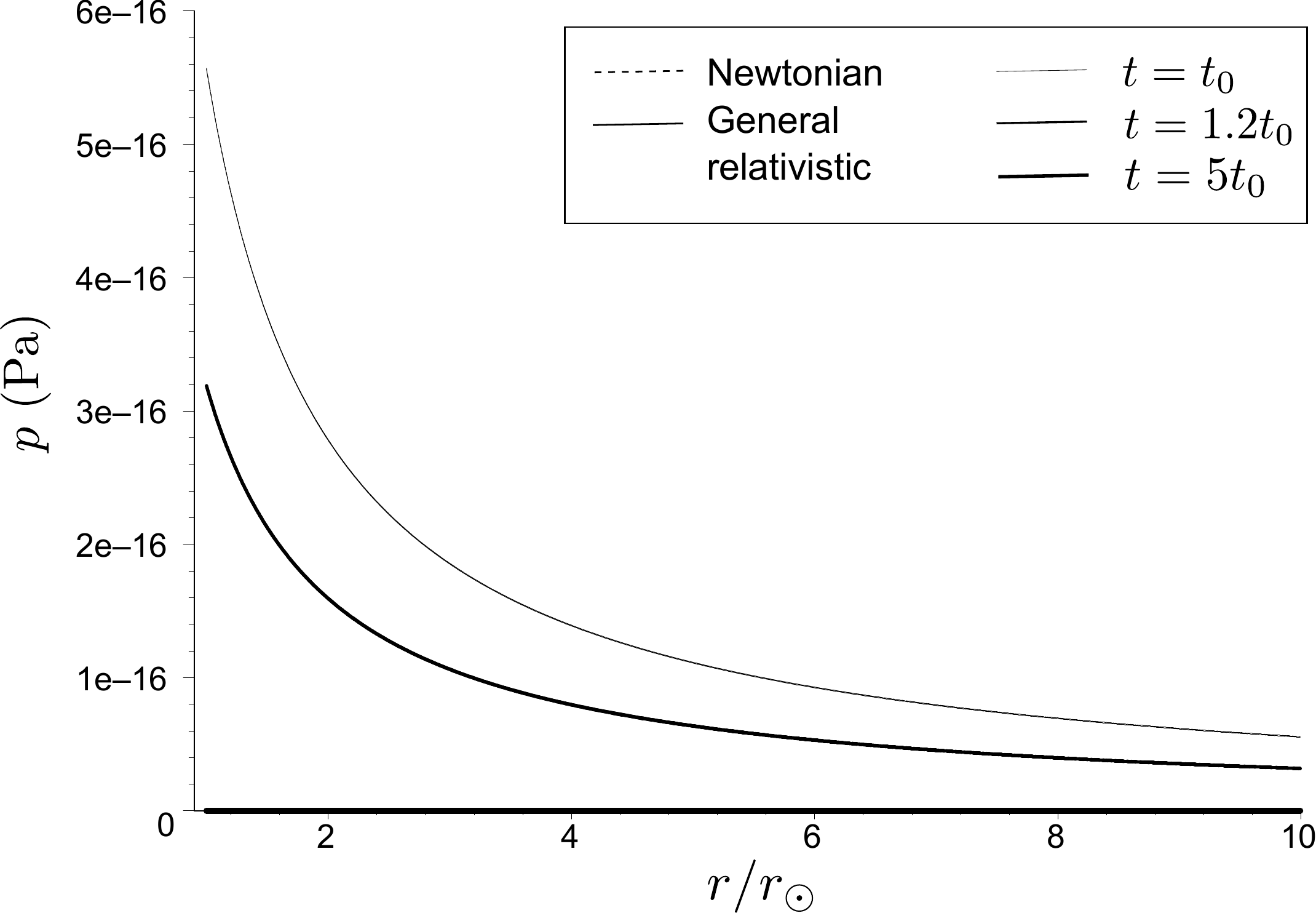}
  \caption{The pressure profile interior (left panel) and exterior
    (right panel) to the Sun, modelled as a uniform-density object
    residing in a concordance $\Lambda$CDM universe, at the present
    cosmological epoch $t=t_0$, and the future times $t=1.2t_0$ and
    $t=5t_0$. For the interior region, the three profiles are
        indistinguishable.  In both regions, the Newtonian and
        general-relativistic results are
        indistinguishable.\label{fig:SunPressure}}
\end{figure*}

In Fig.~\ref{fig:SunPressure}, we plot the resulting interior and
exterior pressure profiles derived, respectively, from the exact
expression (\ref{eq:intpressurestatic}) and approximate result
(\ref{eq:approxpext}), at three different times: the present
cosmological epoch $t=t_0$, for which $H_{\rm e}(t_0) \equiv H_0
= 70$ km s$^{-1}$ Mpc$^{-1}$ \citep{WMAP}, and the future times
$t=1.2t_0$ and $t=5t_0$.
We note that the resulting exterior pressure profiles are
indistinguishable from those obtained by numerically integrating the
exact exterior equation (\ref{eq:eg1OV}). At the level of resolution available in a plot, then the interior profile of course just corresponds to the static part of 
equation (\ref{eq:approx-uplift}), since $p_b(t)$ is so tiny. The 
ratio of the external pressure to typical internal values can be seen by comparing the axes of two plots in Fig.~\ref{fig:SunPressure}, which we can see differ by factors $\approx 10^{29}$. To give some specific figures, at the
    three cosmic times $t=t_0$, $t=1.2t_0$ and $t=5t_0$, the pressure
    at the boundary is $p_{\rm b}(t) \approx 5.6\times 10^{-16}$,
    $3.2\times 10^{-16}$ and $3.0\times 10^{-20}$ Pa, respectively,
    whereas the pressure at the centre of the object is $p_{\rm
      c}(t)\approx 1.33\times 10^{14}$ Pa at all three times. (We note that in
reality, one would expect $p_{\rm c}(t)$ to be larger than this, since
the density of the Sun increases towards the centre rather than being
uniform, as in our model.  For example, in \cite{pain} the central
pressure of the Sun is found to be $\sim 10^{16}$ Pa, which is $\sim
100$ times greater than the value for a uniform density.)

\subsection{Special case of vacuum exterior region}

As a another example, and to make contact with some well-established
results, it is of interest to consider the special case of a vacuum
exterior region, for which $\rho_{\rm e} =0=p_{\rm e}$, but with
$\Lambda \neq 0$, in general. For the purposes of illustration, we
will restrict our attention to a spatially-flat $(k=0)$ exterior
cosmology, for which (\ref{eq:cosmo2}) reduces simply to the standard,
time-independent de-Sitter result $H_{\rm e}^2 =\tfrac{1}{3}\Lambda$.

For a vacuum exterior region, $M = m = {\cal M}=\tfrac{4}{3}\pi\rho_{\rm i}
a^3$ and the $L_tM$ and $L_t\rho$ equations in (\ref{eq:alleqns}) are
satisfied identically, and so we cannot use our earlier general
results for $v_{\rm e}$, $g_{1,{\rm e}}$ and $f_{1,{\rm e}}$, which were derived from
them. Instead, using the $L_rf_1$, $L_rg_1$ and $L_tg_2$ equations in
(\ref{eq:alleqns}) and demanding continuity of $v$ and $g_1$ at the
object boundary, it is straightforward to show that one obtains the
time-independent forms $v_{\rm e} = 0$, $g^2_{1,{\rm e}}(r) = 1-2{\cal M}/r -
\tfrac{1}{3}\Lambda r^2$ and $f_{1,{\rm e}}(r) = 1/g_{1,{\rm e}}(r)$.  Thus, as
one would expect, the corresponding exterior line-element
(\ref{eq:generalmetric}) reduces to the standard, time-independent
(diagonal) Schwarzschild--de-Sitter (SdS) form.

In the interior region, our earlier expressions $v_{\rm i}=0$ and
$g^2_{1,{\rm i}}(r) = 1-2{\cal M}r^2/a^3 -\tfrac{1}{3}\Lambda r^2$ still
hold. To obtain an expression for $f_{1,{\rm i}}$, one first notes that, for
a vacuum exterior, the boundary pressure $p_{\rm b}(t)=0$, from which
one can derive an expression for the central pressure $p_{\rm c}(t)$ from
(\ref{eq:intpressurestatic}).
Subsituting this expression into (\ref{eq:eg1-f1sol}) and noting that,
at the boundary, $f_{\rm 1,b} = 1/\sqrt{1-2{\cal M}/a
  -\tfrac{1}{3}\Lambda a^2}$, one obtains the time-independent result
\begin{widetext}
\begin{equation}
f_{1,{\rm i}}(r)=\frac{2{\cal M}+\tfrac{1}{3}\Lambda a^3}
{3{\cal M}\sqrt{1-\frac{2{\cal M}}{a}-\tfrac{1}{3}\Lambda a^2}-({\cal M}-\tfrac{1}{3}\Lambda a^3)\sqrt{1-\frac{2{\cal M}r^2}{a^3}-\tfrac{1}{3}\Lambda r^2}}.
\end{equation}
\end{widetext}
In the case of a vanishing cosmological constant $\Lambda=0$, the
corresponding line-element (\ref{eq:metric1}) immediately reduces to
that of the well-known Schwarzschild uniform-density interior solution
(see e.g. \cite{GRbook}).

\section{Object with $H_{\rm i}(t)=H_{\rm e}(t)$}\label{sec:nolan}

As our second special case, we now consider an object whose internal
Hubble parameter matches that of the background, so that $H_{\rm
  i}(t)=H_{\rm e}(t)$. In this case, comparing (\ref{eq:g2ext}) and
(\ref{eq:v0elegant}), one immediately finds that
$m(t)=m=\mbox{constant}$, so the excess energy (mass) of the object is
constant, but its total energy $M(\rad(t),t)$ is still, in general,
time-dependent; in this sense this special case is complementary to
the previous one considered in Section~\ref{sec:static}.

\subsection{Exterior region}

In this example, it is more convenient to consider first the exterior
region, for which $M=\tfrac{4}{3}\pi\rho_{\rm e}(t)r^3 + m$, where we
reiterate that $m$ is time-independent. This expression is thus identical
to that originally considered in \cite{NLH1}, except the object
was there taken to be a point mass $m$ and so this form for $M$ was valid
for all positive values of $r$, rather than just for $r >
a(t)$. Nonetheless, the required solutions for $f_{1,{\rm e}}$, $g_{1,{\rm e}}$
and $g_{2,{\rm e}}\equiv v_{\rm e}$ are therefore precisely those listed in Table~1 of
\cite{NLH1}, which also gives a detailed discussion of the appropriate
boundary conditions at large $r$ in the case of a closed $(k=1)$
exterior cosmology.

It is worth noting, however, that our general expressions
(\ref{eq:v0elegant}) and (\ref{eq:g1elegant}) for $v_{\rm e}$ and
$g_{1,{\rm e}}$, respectively, do correctly reduce to the forms
\begin{align}
v_{\rm e} &= rH_{\rm e}(t), \nonumber \\
g_{1,{\rm e}} &= \sqrt{1-\frac{2m}{r}-\frac{kr^2}{R^2(t)}},
\end{align}
as derived in \cite{NLH1}. For our later discussion, it is also worth
reiterating here that, for a spatially-flat $(k=0)$ exterior
cosmology, one obtains $f_{1,{\rm e}}=1/\sqrt{1-2m/r}$, so the
corresponding line-element in the exterior region is given by
\begin{align}
ds^2&=\left[1-\frac{2m}{r}-r^2H_{\rm e}^2(t)\right]dt^2\nonumber \\
&\hspace{1.3cm}+\frac{2rH_{\rm e}(t)} {\sqrt{1-\frac{2m}{r}}}\,dt\,dr
-\frac{1}{1-\frac{2m}{r}}\,dr^2-r^2\,d\Omega^2.\label{eq:flat}
\end{align}
As shown in \cite{NLH1}, by converting to isotropic and comoving
coordinates via $r=\overline{r}R(t)(1+m/(2\overline{r}R(t)))^2$,
this is equivalent to McVittie's line-element
\cite{mcvittie, mcvittie2}, given by
\begin{align}
ds^2=&\left(\frac{1-\frac{m}{2\overline{r}R(t)}}{1+\frac{m}{2\overline{r}R(t)}}\right)^2dt^2\nonumber\\
& \qquad\qquad-\left(1+\frac{m}{2\overline{r}R(t)}\right)^4R^2(t)(d\overline{r}^2+\overline{r}^2d\Omega^2),\label{eq:mcvittieflat}
\end{align}

\subsection{Interior region}

Let us now turn to the interior region, for which
$M=\tfrac{4}{3}\pi\rho_{\rm i}(t)r^3$. Expressions for $v_{\rm i}$ and $g_{1,{\rm i}}$
may be written down immediately by substituting $H_{\rm i}(t)=H_{\rm e}(t)$ into
(\ref{eq:g2int}) and (\ref{eq:g1int}), and
using (\ref{eq:cosmo2}) to simplify; these are given in Table
\ref{table:intparams}. In particular, we see that $v_{\rm i}=rH_{\rm e}(t)=v_{\rm e}$,
so the fluid motion in both the interior and exterior regions is
described by a standard cosmological expansion (for later use, we also
note that one trivially obtains $F_{\rm i}=H_{\rm e}(t)=F_{\rm e}$.)

\begin{table*}
 \caption{Tetrad components defining the interior metric for an object with
   $H_{\rm i}(t)=H_{\rm e}(t)$ in a spatially-flat $(k=0)$, open
  $(k=-1)$ and closed $(k=1)$ expanding universe.\label{table:intparams}}
 \begin{ruledtabular}
 \begin{tabular}{lc}
$f_{1,{\rm i}}^{\{k=0\}}$
& $\frac{2}{3\sqrt{1-\frac{2m}{\rad(t)}}-\sqrt{1-\frac{2mr^2}{\rad^{3}(t)}}}$
\\ [6mm]
$f_{1,{\rm i}}^{\{k=-1\}}$ & $1+\left[\frac{R^2(t)}{\rad^{3}(t)}-\frac{2}{R(t)}\sqrt{1+\frac{r^2}{R^2(t)}}+\left(\frac{1}{\rad(t)}+\frac{2\rad(t)}{R^2(t)}-\frac{R^2(t)}{\rad^{3}(t)}\right)\sqrt{\frac{1+\frac{r^2}{R^2(t)}}{1+\frac{\rad^{2}(t)}{R^2(t)}}}\right]m+O(m^{2})$ \\ [6mm]
$f_{1,{\rm i}}^{\{k=1\}}$ & $1+\left[-\frac{R^2(t)}{\rad^{3}(t)}+\left(\frac{1}{\rad(t)}-\frac{2\rad(t)}{R^2(t)}+\frac{R^2(t)}{\rad^{3}(t)}\right)\sqrt{\frac{1-\frac{r^2}{R^2(t)}}{1-\frac{\rad^{2}(t)}{R^2(t)}}}\right]m+O(m^{2})$\\ [6mm]
$g_{1,{\rm i}}$
& $\sqrt{1-\frac{2mr^2}{\rad^{3}(t)}-\frac{kr^2}{R^2(t)}}$ \\ [4mm]
$g_{2,{\rm i}} \equiv v_{1}$
& $rH_{\rm e}(t)$ \\
\end{tabular}
\end{ruledtabular}
\end{table*}

For the purposes of illustration, we will obtain an expression for
$f_{1,{\rm i}}$ in this case via a slightly different route from that
employing the generalised Oppenheimer--Volkov equation
(\ref{eq:OVgeneralised}), which we used previously. Our previous
approach is still applicable, but for our later analysis it is more
convenient here to obtain $f_{1,{\rm i}}$ by substituting our expressions
for $v_{\rm e}$ and $M$ directly into the expression (\ref{eq:Gsofar}) for
$G_{\rm i}$. We may eliminate $\rho_{\rm i}(t)$ from the resulting expression
in favour of $\rho_{\rm e}(t)$ by using the definition of $m$ to write
\begin{equation}
\rho_{\rm i}(t)=\rho_{\rm e}(t)+\frac{3m}{4\pi \rad^3(t)},\label{eq:rho1rho0rel}
\end{equation}
and, using the results (\ref{eq:cosmo1})--(\ref{eq:cosmo3}) and
(\ref{eq:lta}), one then obtains
\begin{equation}
G_{\rm i} =  \left(\frac{r}{R^2(t)}\right)\frac{k(f_{1,{\rm i}}-1)+\frac{mR^2(t)}{a^3(t)}
\left(\frac{3f_{1,{\rm i}}}{f_{1,b}(t)}-2\right)}{\sqrt{1-\frac{2mr^2}{a^3(t)}-\frac{kr^2}{R^2(t)}}}.
\label{eq:G1sofar}
\end{equation}
Finally, one may obtain $f_{1,{\rm i}}$ by substituting $G_{\rm i}$ into the
$L_rf_1$-equation in (\ref{eq:alleqns}) and integrating. To perform
this integral, however, one requires an expression for $f_{1,b}(t)$,
and, moreover, the resulting constant of integration can only be
determined by the imposition of suitable conditions at the object
boundary. Thus the solution for $f_{1,{\rm i}}$ depends on the nature of the
exterior cosmology. We now consider each case individually.

\subsection{Interior metric for a spatially-flat background}

As mentioned above, for a spatially-flat ($k=0$) exterior cosmology
one has $f_{1,{\rm e}}^{\{k=0\}}=1/\sqrt{1-2m/r}$, from which one obtains
the boundary condition
\begin{equation}
f_{1,{\rm b}}^{\{k=0\}}(t)=\frac{1}{\sqrt{1-\frac{2m}{\rad(t)}}}.\label{eq:flatbound}
\end{equation}
Using the $L_rf_1$-equation in (\ref{eq:alleqns}), one then finds that
$f_{1,{\rm i}}^{\{k=0\}}$ must satisfy the differential equation
\begin{equation}
\frac{1}{f_{1,{\rm i}}}\frac{df_{1,{\rm i}}}{dr}+\frac{mr}{\rad^{3}(t)\left(1-\frac{2mr^2}{\rad^{3}(t)}\right)}\left(3f_{1,{\rm i}}\sqrt{1-\frac{2m}{\rad(t)}}-2\right)=0,\nonumber
\end{equation}
where, for brevity, we have omitted the superscript relating to a
spatially-flat background.  This equation may be integrated to obtain
the analytical solution
\begin{equation}
f_{1,{\rm i}}^{\{k=0\}}=\frac{2}{3\sqrt{1-\frac{2m}{\rad(t)}}+2C(t)\rad^{3/2}(t)\sqrt{1-\frac{2mr^2}{\rad^{3}(t)}}},\nonumber
\end{equation}
where, using the boundary condition (\ref{eq:flatbound}), the
integration constant is found to be
$C(t)=-(1/2)\rad^{-3/2}(t)$.  Thus, the final solution for
$f_{1,{\rm i}}^{\{k=0\}}$ is as stated in Table \ref{table:intparams}, and
the interior line-element can be written explicitly using equation
(\ref{eq:generalmetric}) as
\begin{widetext}
\begin{align}
ds^2=\frac{\left[1-\frac{2mr^2}{\rad^{3}(t)}-r^2H_{\rm e}^2(t)\right]\left[3\sqrt{1-\frac{2m}{\rad(t)}}-\sqrt{1-\frac{2mr^2}{\rad^{3}(t)}}\right]^2}{4\left(1-\frac{2mr^2}{\rad^{3}(t)}\right)}dt^2 &+\frac{rH_{\rm e}(t)\left[3\sqrt{1-\frac{2m}{\rad(t)}}-\sqrt{1-\frac{2mr^2}{\rad^{3}(t)}}\right]}{\left(1-\frac{2mr^2}{\rad^{3}(t)}\right)}\,dt\,dr\nonumber\\
 &-\frac{1}{1-\frac{2mr^2}{\rad^{3}(t)}}dr^2
-r^2d\Omega^2.\label{eq:interiormetric}
\end{align}
\end{widetext}

We note that, in the case where there is no expansion,
$H_{\rm i}(t)=H_{\rm e}(t)=0$, so that the boundary of the object remains fixed,
$\rad(t)=\rad$, the above result correctly reduces to the well-known
Schwarzschild interior solution for a uniform density object (see
e.g. \cite{GRbook}). We now investigate some important physical
properties of our newly-derived interior metric.

\subsubsection{Densities and pressures}\label{sec:densandpress}

Let us first consider the density and pressure of the interior fluid.
Using the general expressions in (\ref{eq:generalpandrho}), and
simplifying using (\ref{eq:lta}) and (\ref{eq:flatbound}), one
obtains
\begin{align}
8\pi\rho_{\rm i}(t)&=3\left(H_{\rm e}^2(t)-\tfrac{1}{3}\Lambda+\frac{2m}{\rad^{3}(t)}\right),  \label{eq:intdensitypressure} \\
8\pi p_{\rm i}&=-3\left(H_{\rm e}^2(t)-\tfrac{1}{3}\Lambda+\frac{2m}{\rad^{3}(t)}\right)\nonumber\\
&\indent -\frac{4\left(H_{\rm e}^{\prime}(t)-\frac{3m}{\rad^{3}(t)}\sqrt{1-\frac{2m}{\rad(t)}}\right)}{3\sqrt{1-\frac{2m}{\rad(t)}}-\sqrt{1-\frac{2mr^2}{\rad^{3}(t)}}}.\label{eq:flatintpressure}
\end{align}
The expression for $\rho_{\rm i}(t)$ could alternatively be obtained by
combining the relationships (\ref{eq:cosmo2}) and
(\ref{eq:rho1rho0rel}).

According to our boundary conditions, the internal pressure profile
must be continuous with the exterior pressure profile at the object
boundary. One may easily show (see, e.g., \citep{NLH1}) that the
latter is given by
\begin{equation}
8\pi p_{\rm e}=-3(H^2_{\rm e}(t)-\tfrac{1}{3}\Lambda)-\frac{2H^{\prime}_{\rm e}(t)}{\sqrt{1-\frac{2m}{r}}},
\end{equation}
which is clearly continuous with (\ref{eq:flatintpressure}) at
$r=\rad(t)$. One may show, however, that the radial derivative of the
pressure profile has a jump there, given by
\begin{equation}
8\pi\frac{\partial p_{\rm e}}{\partial r}\Big |_{r=\rad(t)}-8\pi\frac{\partial
p_{\rm i}}{\partial r}\Big |_{r=\rad(t)}=\frac{6m^2}{\rad^{5}(t)\left(1-\frac{2m}{\rad(t)}\right)}.
\end{equation}

\begin{figure*}
\centering
\begin{tabular}{ccc}
\begin{minipage}{2.2in}
\centering
\includegraphics[height=2in,width=2.2in]{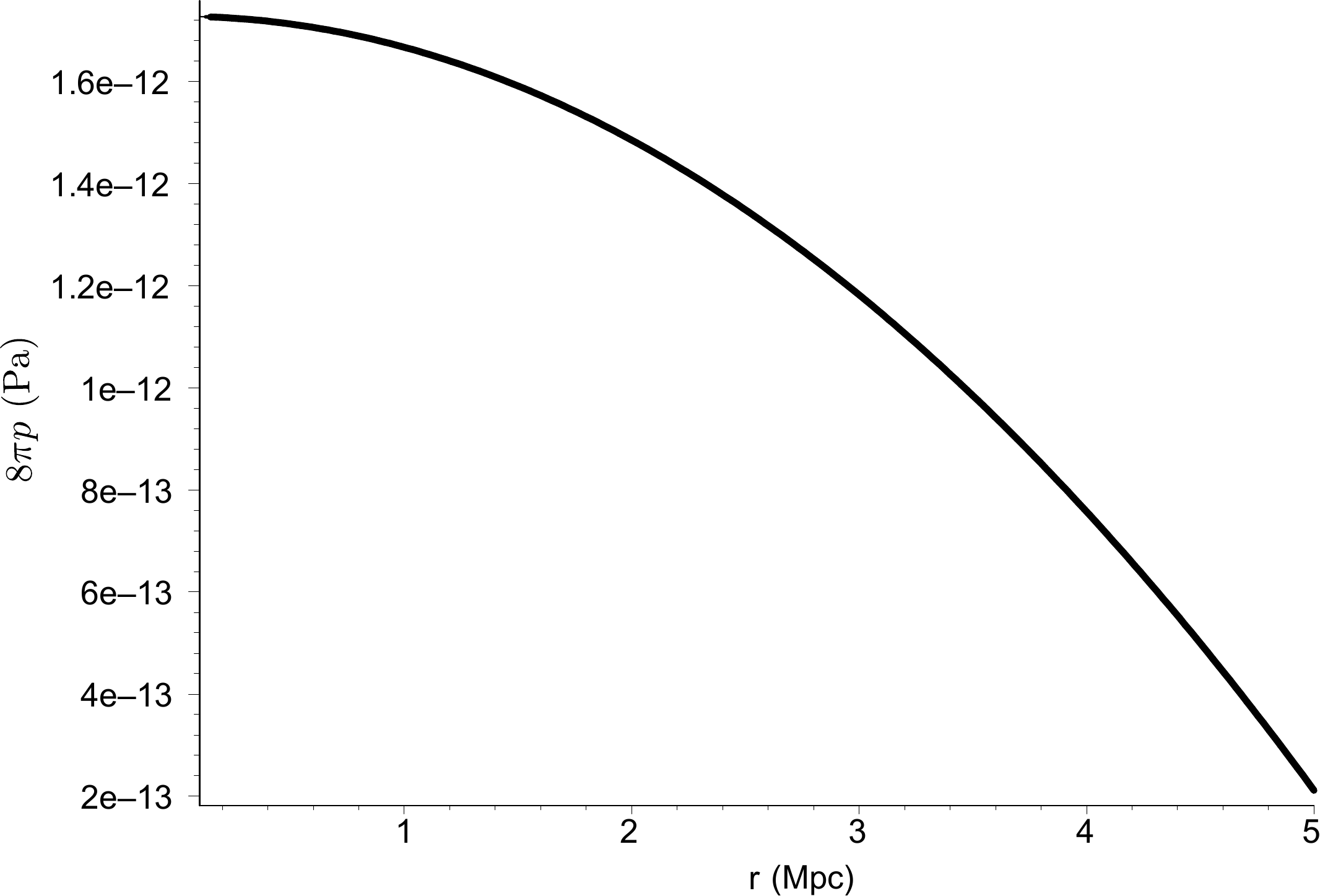}
\end{minipage}
\hspace{0.01cm}
&
\begin{minipage}{2.2in}
\centering
\includegraphics[height=2in,width=2.2in]{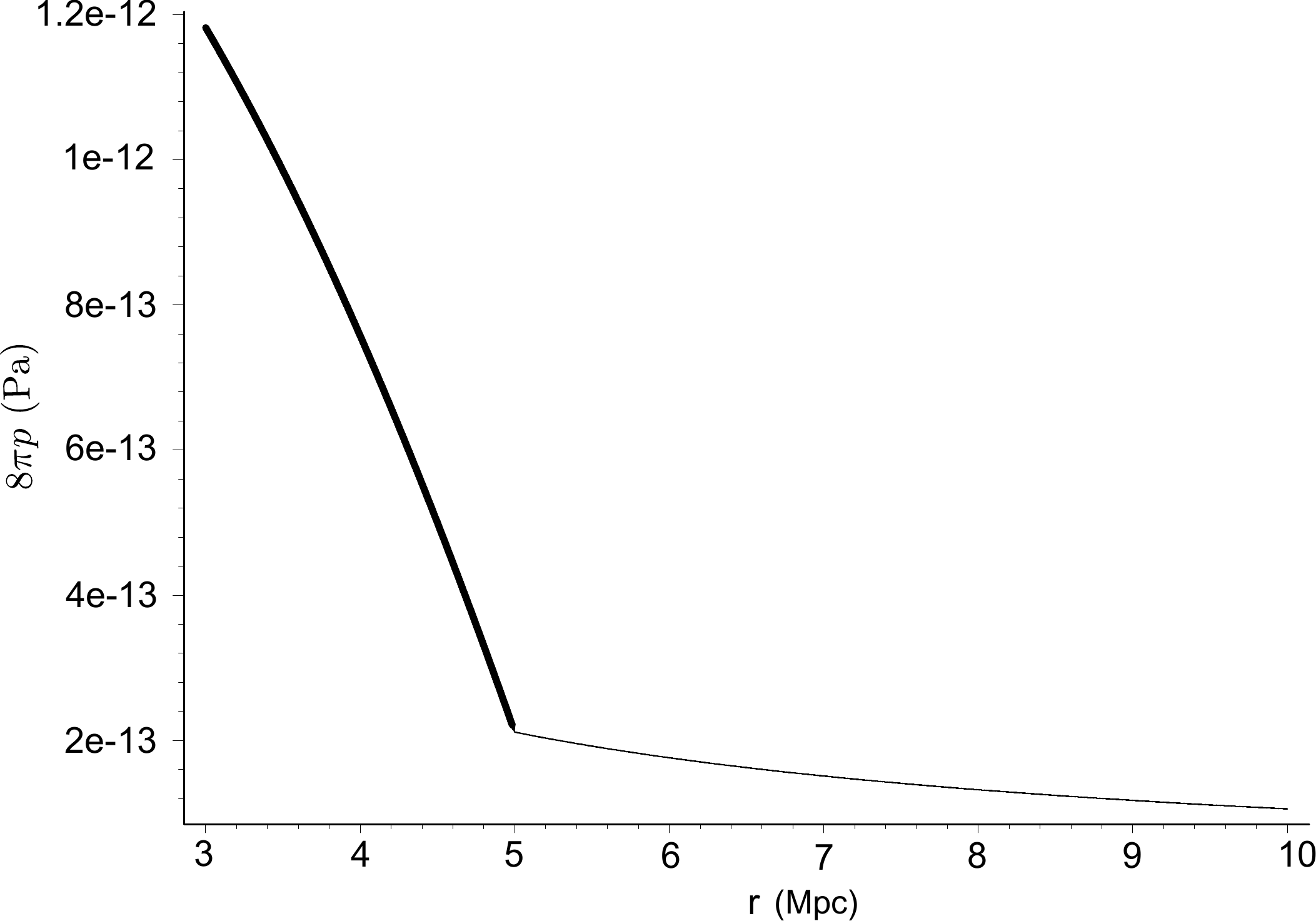}
\end{minipage}
&
\begin{minipage}{2.2in}
\centering
\includegraphics[height=2in,width=2.2in]{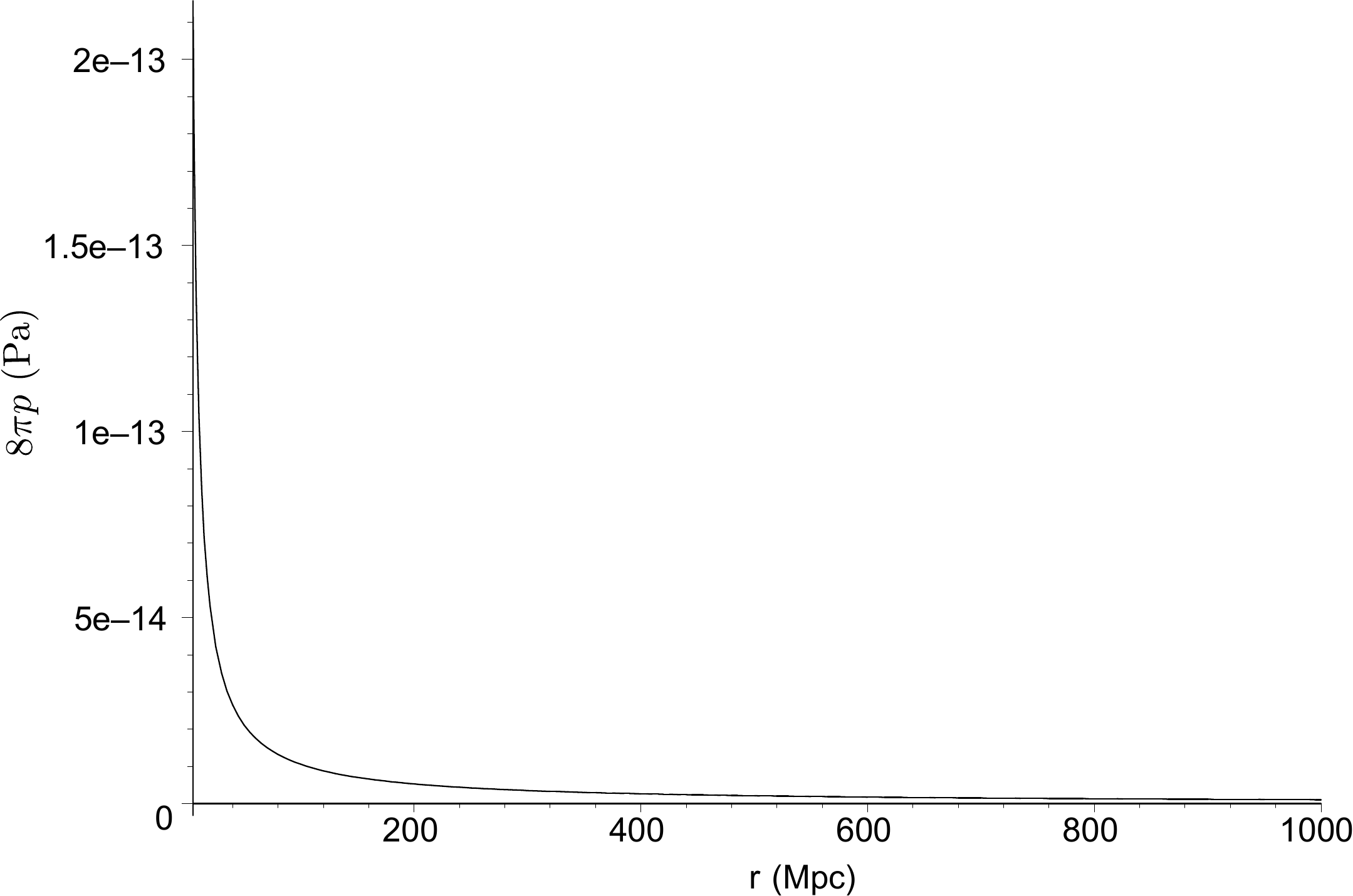}
\end{minipage}
\hspace{0.01cm}
\\
\\
(a) $p_{\rm i}$ &
(b) $p_{\rm i}$  and $p_{\rm e}$ &
(c) $p_{\rm e}$
\end{tabular}
\caption{The radial pressure profile for an approximate model of
    cluster of galaxies, taken to be a uniform-density object with
    excess mass $m=10^{15} M_{\odot}$ and current radius
    $\rad(t_0)=5$~Mpc, for (a) $r < 5$~Mpc; (b) $3< r < 10$~Mpc; and
    (c) $5 < r < 1000$~Mpc. The exterior cosmology is assumed to be
    concordance $\Lambda$CDM. \label{fig:pressure}}
\end{figure*}

As an illustration of these results, we show in
Fig.~\ref{fig:pressure} the complete pressure profile for an object
intended to approximate a cluster of galaxies, here modelled as a
uniform-density sphere with excess mass $m=10^{15} M_{\odot}$ and
current radius $\rad(t_0)=5$~Mpc. This is embedded in a concordance
$\Lambda$CDM exterior cosmology, for which $H_e(t)$ is given by
(\ref{eq:LambdaCDM}), and assuming $p_{\infty}=0$. The discontinuity
in the radial pressure gradient at $r=5{\rm \, Mpc}$ can be clearly
seen.

Although we expect the pressure to be finite at the centre of the
object, one can see from (\ref{eq:flatintpressure}) that it, in fact,
diverges when $3\sqrt{1-2m/\rad(t)}=1$.  This places an upper bound on
the mass/size ratio of any object
\begin{equation}
\frac{m}{\rad(t)}<\frac{4}{9}.\label{eq:buchdahl}
\end{equation}
This is equivalent to Buchdahl's theorem \citep{GRbook} for the
Schwarzschild uniform-density interior metric, but with
$\rad\rightarrow \rad(t)$.

\subsubsection{Evolution of the object boundary}\label{sec:physicalparams}

We now consider the rate of expansion of the object boundary. From
(\ref{eq:lta}) and (\ref{eq:flatbound}), one finds
\begin{equation}
\rad^{\prime}(t)=H_{\rm e}(t)\rad(t)\sqrt{1-\frac{2m}{\rad(t)}}.\label{eq:diffr1flat}
\end{equation}
Although elegant, this result indicates the unrealistic nature of this
model for the description of compact astrophysical objects.  Since
$m/\rad(t)\ll 1$ for most such objects, the above result suggests that
they expand more or less with the Hubble flow, which is clearly
unrealistic. For example, both the Sun (a compact object), which has
mass $m\approx 2\times10^{30}$~kg and current size $\rad(t_0) \approx
7\times 10^8$~m, and a galaxy cluster with mass $\sim
10^{15}M_{\odot}$ and size $\rad(t_0) \sim 5$~Mpc, have
$m/\rad(t_0)\approx 10^{-6}$, and are therefore within this model
predicted to be currently expanding at very close to the Hubble rate.
While this might be plausible for a cluster, it clearly does not apply
to the Sun.  Thus, the expected resistance of more compact objects to
the external Hubble flow is not encapsulated in the expression
(\ref{eq:diffr1flat}), and results from our initial assumption that
$H_{\rm i}(t)=H_{\rm e}(t)$.

Integrating (\ref{eq:diffr1flat}) leads directly to
\begin{equation}
\rad(t)=\bar{a}R(t)\left(1+\frac{m}{2\bar{a}R(t)}\right)^2,
\label{eq:r1t}
\end{equation}
where $\bar{a}$ is a dimensionless constant.
Solving this equation for $\bar{a}$ yields the two roots
\begin{equation}
\bar{a}=\frac{\rad(t)}{2R(t)}\left[\left(1-\frac{m}{\rad(t)}\right)\pm\sqrt{1-\frac{2m}{\rad(t)}}\right].\label{eq:roverbar0}
\end{equation}
From (\ref{eq:buchdahl}), we require $m/\rad(t)<4/9$, in which case
both roots lead to positive values for $\bar{a}$, as one would expect.
Indeed, combining (\ref{eq:buchdahl}) and (\ref{eq:r1t}), one obtains
the condition
\begin{equation}
m^2-5R(t)\bar{a} m+4R^2(t)\rad^{2}>0,\nonumber
\end{equation}
which factorises and leads to the following
bounds on the ratio $m/\bar{a}$ of any object in this model:
\begin{equation}
\frac{m}{\bar{a}}<R(t)\qquad\text{or}\qquad
\frac{m}{\bar{a}}>4R(t),
\end{equation}
where the former bound applies to the plus sign in
(\ref{eq:roverbar0}), and the latter bound to the minus sign.  Since
one is usually interested in objects for which $m\ll \rad(t)$, for
which the former bound holds, the plus sign is the relevant one in
(\ref{eq:roverbar0}). Hence, if one wishes to determine the radius of
the object boundary at some time $t$ using equation (\ref{eq:r1t}), one
first calculates $\bar{a}$ by substituting the object's current
radius $\rad(t_0)$, its excess mass $m$ and the current exterior scale
factor $R(t_0)$ into (\ref{eq:roverbar0}) (using the plus sign).

In addition to considering the evolution of the object coordinate
radius $a(t)$, it is also of interest to consider the variation with
time of the proper distance $l(t)$ from the centre to the boundary of
the object.  From the line-element (\ref{eq:interiormetric}), one sees
that the proper distance is, in general, defined through
$dl=\sqrt{1-\frac{2mr^2}{\rad^{3}(t)}}dr$.  Therefore, $l(t)$ is
found by integrating this expression between $r=0$ and $r=a(t)$, which gives
\begin{align}
l(t)&=\frac{1}{\sqrt{2m/\rad^{3}(t)}}\arcsin\sqrt{\frac{2m}{\rad(t)}},\nonumber\\
& \approx \rad(t)+\tfrac{1}{3}m,\label{eq:l1}
\end{align}
where in the second line we have taken the limit $m\ll \rad(t)$, such
that the Schwarzschild radius of the object lies well within its
boundary. One can therefore see that the proper distance from the
centre to the boundary of the object is only slightly greater than its
coordinate radius.

\subsubsection{Comparison with Nolan's interior metric}

In \cite{nolan}, Nolan also derived an interior metric for an object
with uniform spatial density residing in a spatially-flat expanding
universe.  For consistency with McVittie's earlier analysis, however,
Nolan worked in isotropic and comoving coordinates; we denote his
dimensionless, comoving radial coordinate by $\bar{r}$. Indeed, in
deriving his metric, Nolan demanded that it matched onto McVittie's
exterior metric at the boundary of the object, which he assumed to be
at a {\em fixed} comoving coordinate radius $\bar{r}=\bar{a}$.  We
note that this assumption clearly implies that the object is
expanding, although Nolan makes no explicit mention of this.

The resulting interior metric reads
\begin{align}
ds^2&=\left[\frac{1-\frac{m}{\bar{a}R(t)}+\frac{m\bar{r}^2}{\bar{a}^{3}R(t)}\left(1-\frac{m}{4\bar{a}R(t)}\right)}{\left(1+\frac{m}{2\bar{a}R(t)}\right)\left(1+\frac{m\bar{r}^2}{2\bar{a}^{3}R(t)}\right)}\right]^2dt^2\nonumber\\
&\qquad\qquad -R^2(t)\frac{\left(1+\frac{m}{2\bar{a}R(t)}\right)^6}{\left(1+\frac{m\bar{r}^2}{2\bar{a}^{3}R(t)}\right)^2}(d\bar{r}^2+\bar{r}^2d\Omega^2),\label{eq:Nolaninterior}
\end{align}
which may be shown to be equivalent to our interior metric
(\ref{eq:interiormetric}) via the coordinate transformation
\begin{equation}
r=\bar{r}R(t)\frac{\left(1+\frac{m}{2\bar{a}R(t)}\right)^3}{1+\frac{m\bar{r}^2}{2\bar{a}^{3}R(t)}}.\label{eq:nolancoordtrans}
\end{equation}
In particular, we note that at the edge of the object
$\bar{r}=\bar{a}$, one recovers our solution (\ref{eq:r1t}) for the
growth of the boundary. The above transformation also relates the
density and pressure associated with Nolan's interior metric to our
expressions in equations (\ref{eq:intdensitypressure}) and
(\ref{eq:flatintpressure}), confirming that Nolan's results correspond
to the subcase $H_i(t)=H_e(t)$ within our set of spatially-flat
models.

\subsection{Interior metric for spatially-curved backgrounds}

One may find the interior metric in the case of a spatially-curved
exterior cosmology using the same methodology as in the
spatially-flat case.  The expressions for $g_{1,{\rm i}}$ and $v_{\rm i}$ are already
known, and given in Table \ref{table:intparams}, so it again only remains
to obtain a form for $f_{1,{\rm i}}$.

We must find a solution for $f_{1,{\rm i}}$ that matches onto the
corresponding exterior solution $f_{1,{\rm e}}$ at $r=\rad(t)$. As shown in
\cite{NLH1}, in a spatially-curved universe $f_{1,{\rm e}}$ is given in
terms of an elliptic integral. To avoid this complication, we
therefore focussed on the region $m\ll r\ll R(t)$, within which we
could represent $f_{1,{\rm e}}$ as a truncated power series in $m$. The
resulting solutions have the form
\begin{equation}
f_{1,{\rm e}}^{\{k=\pm1\}}=b_{0,{\rm e}}+b_{1,{\rm e}} m+O(m^2),\nonumber
\end{equation}
where, for each type of universe, the coefficients $b_{n,{\rm e}}$ were found to be
\begin{align}
b_{0,{\rm e}}^{\{k=\pm1\}}&=1,\nonumber\\
b_{1,{\rm e}}^{\{k=-1\}}&=\frac{1}{r}+\frac{2r}{R^2(t)}-\frac{2}{R(t)}\sqrt{1+
\frac{r^2}{R^2(t)}},\nonumber\\
b_{1,{\rm e}}^{\{k=1\}}&=\frac{1}{r}-\frac{2r}{R^2(t)}.\label{eq:f1NLH1}
\end{align}

\begin{figure*}
\centering
\begin{tabular}{cc}
\begin{minipage}{3in}
\centering
\includegraphics[height=2in,width=2.2in]
{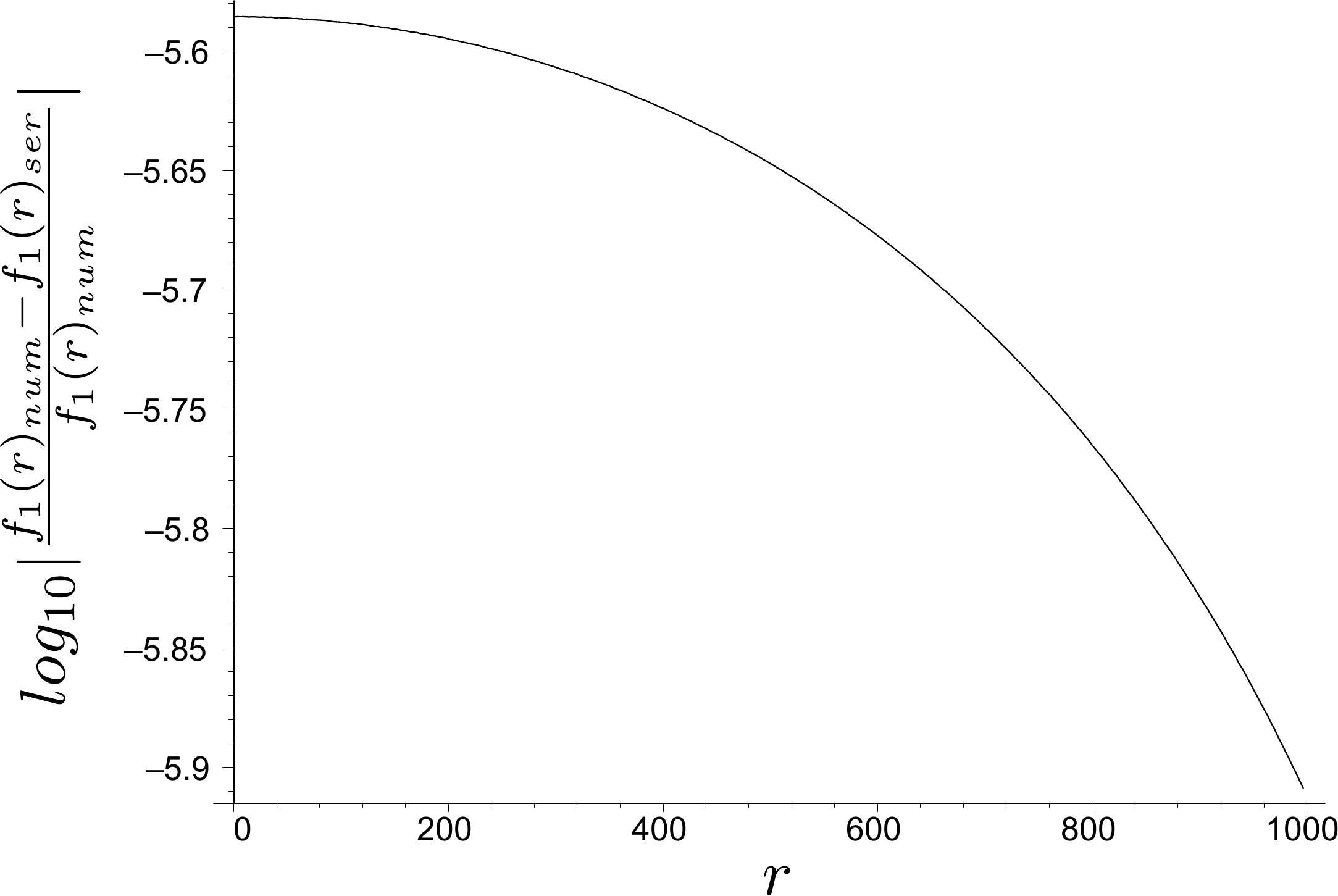}
\end{minipage}
&
\begin{minipage}{3in}
\centering
\includegraphics[height=2in,width=2.2in]
{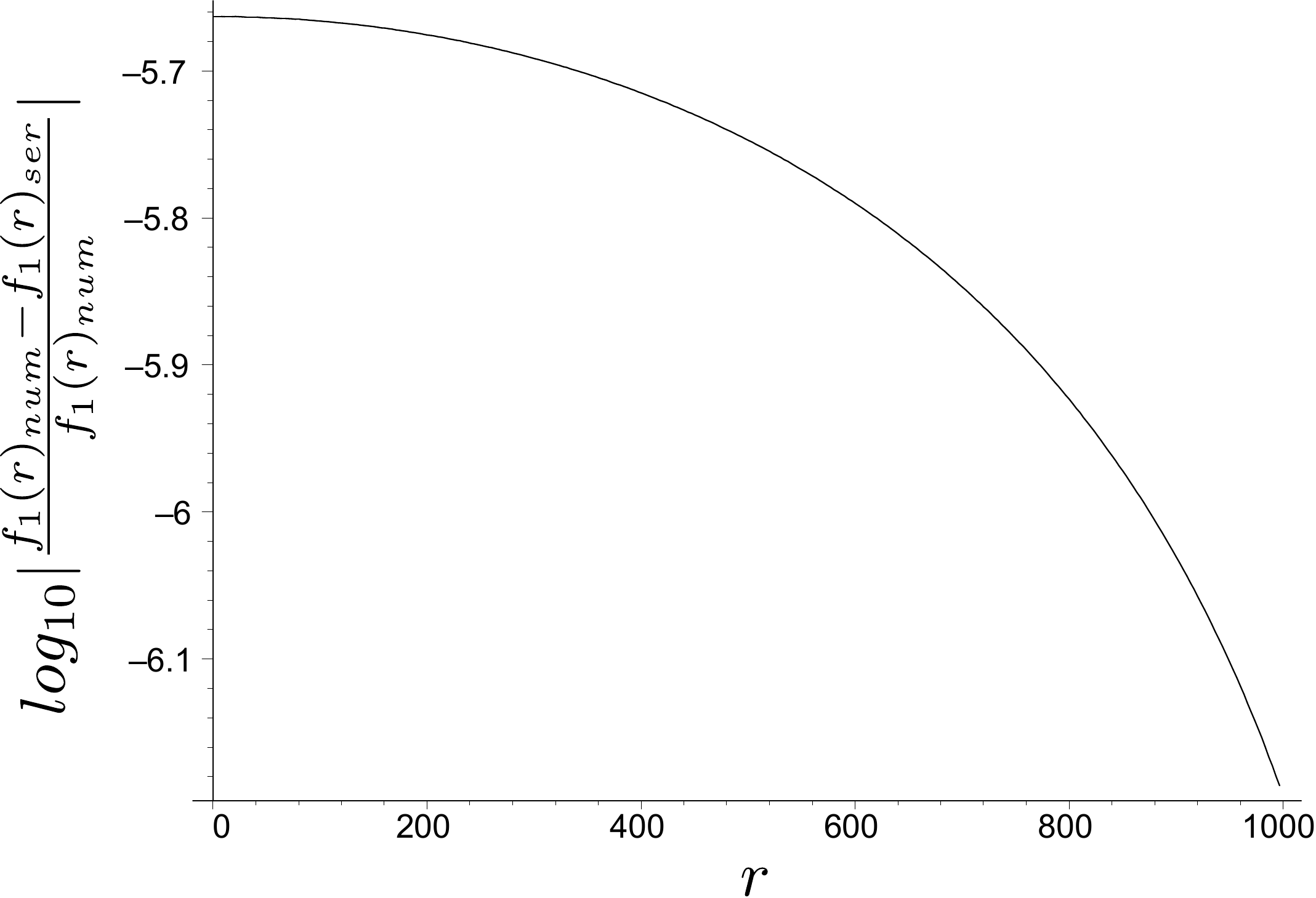}
\end{minipage}
\\
\newline
\\
(a) $k=-1$ &
(b) $k=1$
\end{tabular}
\caption{The logarithm of the fractional error in the approximate
  series solution for the interior quantity $f_{1,{\rm i}}$ (Table
  \ref{table:intparams}), relative to the exact numerical result, for
  fixed arbitrary values $m=1$, $\rad(t)=1000$ and
  $R(t)=10000$.}\label{fig:f1comparison}
\end{figure*}

We now solve for $f_{1,{\rm i}}^{\{k=\pm1\}}$, also for $m\ll r\ll R(t)$, by
looking for a power series solution of the form
$f_{1,{\rm i}}^{\{k=\pm1\}}=b_{0,{\rm i}}+b_{1,{\rm i}} m+O(m^2)$, whose coefficients
match the exterior solution at the boundary of the object.  We begin
by substituting the appropriate expressions for $G_{1}$ and $g_{1,{\rm i}}$,
given in (\ref{eq:G1sofar}) and Table \ref{table:intparams},
respectively, into the $L_r f_1$-equation in (\ref{eq:alleqns}).
The resulting differential equation may be integrated
without specifying a value of $k$, to obtain
\begin{align}
\frac{1}{f_{1,{\rm i}}^{\{k=\pm1\}}}&=\frac{k\rad^{3}+3mR^2/f_{1,b}^{\{k=\pm1\}}}{k\rad^{3}+2mR^2}\nonumber\\
&\indent\indent+CR\rad^{3/2}\sqrt{1-\frac{2mr^2}{\rad^{3}}-\frac{kr^2}{R^2}},\label{eq:f1curvedsofar}
\end{align}
where, for brevity, we have momentarily dropped the dependencies on
$t$, and $C(t)$ arises as a constant of integration.
Evaluating this relationship at $r=\rad(t)$ allows one to solve for
$f_{1,b}^{\{k=\pm1\}}(t)$ in terms $C(t)$:
\begin{equation}
f_{1,b}^{\{k=\pm1\}}(t)=\frac{\frac{k\rad^{2}}{R^2}-\frac{m}{\rad}}{\frac{k\rad^{2}}{R^2}+CR\rad^{3/2}\sqrt{1-\frac{2m}{\rad}-\frac{k\rad^{2}}{R^2}}\left(\frac{k\rad^{2}}{R^2}+\frac{2m}{\rad}\right)}.\nonumber
\end{equation}
The series expansion of this expression in $m$ should equal the
corresponding exterior result.  Before expanding, however, it is
first necessary also to express the function $C(t)$ as a power
series in $m$.  Considering $C(t)=c_0+c_1m+O(m^{2})$, the
coefficients $c_n$ then appear embedded within the interior
coefficients $b_{n,{\rm i}}$.  In order to satisfy
$b_{n,{\rm e}}(\rad(t),t)=b_{n,{\rm i}}(\rad(t),t)$ it can be shown that
\begin{align}
c_{1}^{\{k=-1\}}&=\frac{R(t)\left(1-\frac{\rad^{2}(t)}{R^2(t)}-\frac{2\rad^{4}(t)}{R^{4}(t)}\right)}{\rad^{9/2}(t)\sqrt{1+\frac{\rad^{2}(t)}{R^2(t)}}}+\frac{2}{\rad^{3/2}(t)R^2(t)},\nonumber \\
c_{1}^{\{k=+1\}}&=\frac{R(t)\left(\frac{2\rad^{4}(t)}{R^4(t)}-\frac{\rad^{2}(t)}{R^2(t)}-1\right)}{\rad^{9/2}(t)\sqrt{1-\frac{\rad^{2}(t)}{R^2(t)}}}.\nonumber
\end{align}
With these solutions we are finally able to obtain expressions for
$b_{n,{\rm i}}$, and hence approximate solutions for $f_{1,{\rm i}}^{\{k=\pm1\}}$,
which are given in Table \ref{table:intparams}.
Fig. \ref{fig:f1comparison} shows that within the object our
approximate solutions for $f_{1,{\rm i}}^{\{k=\pm1\}}$ are sufficient to
represent the true solutions to high accuracy.

It is now possible to write down metrics describing the
spacetime interior to an object with constant spatial density residing
in an open ($k=-1$) or closed ($k=1$) expanding universe.  From
equation (\ref{eq:generalmetric}) they are given by
\begin{widetext}
\begin{equation}
ds^2 = \frac{1-\frac{2mr^2}{\rad^{3}(t)}-\frac{kr^2}{R^2(t)}-\rad^{2}(t)H^2(t)}{f_{1,{\rm i}}^{2}\left(1-\frac{2mr^2}{\rad^{3}(t)}-\frac{kr^2}{R^2(t)}\right)}dt^2
 +\frac{2rH(t)}{f_{1,{\rm i}}\left(1-\frac{2mr^2}{\rad^{3}(t)}-\frac{kr^2}{R^2(t)}\right)}drdt -\frac{1}{1-\frac{2mr^2}{\rad^{3}(t)}-\frac{kr^2}{R^2(t)}}dr^2-r^2d\Omega^2,\label{eq:interiormetriccurved}
\end{equation}
\end{widetext}
where the functions $f_{1,{\rm i}}=f_{1,{\rm i}}^{\{k=\pm1\}}$ are stated in Table \ref{table:intparams}.

\subsection{Force on a test particle}

As our final investigation into the physical properties of our
newly-derived interior metrics, we now consider the force experienced
by a test particle at various locations in our model system. By
considering an idealised infinitesimal test particle, we exclude
forces due to pressure gradients in the fluid and focus only on the
gravitational force experienced. In particular, we derive the force
that must be exerted on a test particle for it to maintain a
particular state of motion in two different circumstances: to remain
at rest, relative to the origin, in the interior of the object; and to
remain on the, in general, moving boundary of the object.

\subsubsection{Particle at rest in the object's interior}

In \cite{NLH1} we derived a fully general-relativistic invariant
expression for the outwards radial force $f$ (per unit rest mass)
required to hold a test particle in any spherically-symmetric
spacetime at rest, i.e. at constant physical radius $r$; this is
equivalent to $\dot{r}=0$, where a dot denotes differentiation with
respect to the particle's proper time. The general expression is given
by
\begin{equation}
f=\frac{1}{\sqrt{g_1^{2}-g_2^{2}}}\left[\frac{f_1(g_2\partial_t g_1-g_1\partial_t g_2)}{g_1^{2}-g_2^{2}}+Gg_1-Fg_2\right].\label{eq:force}
\end{equation}
We note that this expression becomes singular at any point where
$g_1^{2}=g_2^{2}$, which be can shown to correspond to the locations
of apparent horizons \cite{FMN}. Note that these do not usually
coincide with where the metric and fluid pressure become singular,
which occurs when $g_1=0$.  We now present approximate expressions for
the force $f$ in our newly-derived interior metrics in the region
$m\ll r\ll a(t)\ll 1/H_{\rm e}(t)$.  This represents our region of interest,
which is deeply embedded within the object but still well outside its
Schwarzschild radius.


In all cases, expanding equation (\ref{eq:force}) in small quantities
and keeping all terms up to second order, leads to the expression
\begin{align}
f &=\frac{mr}{\rad^{3}(t)}+\frac{3}{2}\frac{m^2r}{\rad^{4}(t)}+q_{\rm e}(t)H_{\rm e}^2(t)r\nonumber\\
&\indent +\frac{3}{2}(1+q_{\rm e}(t))\frac{mH_{\rm e}^2(t)r}{\rad(t)}+\frac{3k}{2}\frac{mr}{\rad(t)R^2(t)},\label{eq:curvedforcelong}
\end{align}
where $q_{\rm e}(t)=-H^{\prime}_{\rm e}(t)/H_{\rm e}^2(t)-1$ is the
deceleration parameter, which currently has a value $q_{\rm e}(t_0)
\equiv q_0 \sim -0.55$ \cite{WMAP}.  One sees that a difference
between the spatially-flat and curved backgrounds occurs only in the
final term, which is of second-order in small quantities and resembles
a type of gravitational force dependent on both the particle's
position and the curvature scale of the universe.

Keeping only terms to first-order in small quantities, one
finds that in all cases
\begin{equation}
f \approx \frac{mr}{\rad^{3}(t)}+q_{\rm e}(t)H_{\rm e}^2(t)r.\label{eq:fintflatapprox}
\end{equation}
At the boundary of the object, this is consistent with the force in
the exterior region obtained in \cite{NLH1}.  Qualitatively, we see that the
force {\it{on}} an interior particle, given by $-f$, comprises an
inwards pointing component due to the mass interior to the particle
and an outwards (inwards) pointing component due to the acceleration
(deceleration) of the background's expansion.

\subsubsection{Particle fixed on the object boundary}

We now consider the force required to keep a test particle located on
the boundary of the object. Since this boundary is moving, one
obviously cannot use the result (\ref{eq:force}), and so we must
derive the appropriate expression.

It was shown in \cite{NLH1} that the force per unit mass required to keep a
radially-moving test particle in some state of motion may be written
in terms of its rapidity $\psi(\tau)$ in the local tetrad frame as
\begin{equation}
\hat{f} = \dot{\psi}(\tau) + G\cosh\psi(\tau) + F\sinh\psi(\tau),
\label{eq:forcegen}
\end{equation}
where dots denote derivatives with respect to the particle's proper
time $\tau$.  Note that we denote this general force $\hat{f}$, so as
not to confuse it with the specific force $f$ required to keep a particle
at rest.  The non-zero components of the particle's four-velocity
in the coordinate basis are
\begin{align}
\dot{t}&=f_1\cosh\psi(\tau),\nonumber\\
\dot{r}&=v\cosh\psi(\tau)+g_1\sinh\psi(\tau).\label{eq:rdotgeneral}
\end{align}

A particle fixed to the boundary of the object has
$\dot{r}=\dot{a}(t)=\rad^{\prime}(t)\dot{t}$, and so the equations
(\ref{eq:rdotgeneral}) yield
$\sinh\psi(\tau)=(\cosh\psi(\tau)/g_1)\left(f_1
\rad^{\prime}(t)-v\right)$, where $g_1$, $v$ and $f_1$ are evaluated at
$r=\rad(t)$.  Using $\cosh^2\psi(\tau)-\sinh^2\psi(\tau)=1$, one obtains
\begin{align}
\cosh\psi(\tau)&=\frac{g_1}{\sqrt{g_1^{2}-x^2}}\Bigg |_{r=\rad(t)},\nonumber\\
\sinh\psi(\tau)&=-\frac{x}{\sqrt{g_1^{2}-x^2}}\Bigg |_{r=\rad(t)},\label{eq:coshsinhint}
\end{align}
where $x=v-f_1\rad^{\prime}(t)$.  Substituting these expressions into
(\ref{eq:forcegen}) then leads to a general expression for
the force required to keep a test particle on the object boundary,
which we denote by $f_{\rad(t)}$. Using
$d/d\tau=\dot{t}\partial_t+\dot{r}\partial_r$ this may be written
(dropping explicit $t$-dependencies for brevity)
\begin{widetext}
\begin{equation}
f_{\rad(t)}=\frac{1}{\sqrt{g_1^{2}-x^2}}
\left\{f_1g_1\left[\frac{x\left(\partial_t g_1+\rad^{\prime}\partial_r g_1\right)-g_1\left(\partial_t v+\rad^{\prime}\partial_r v\right)}{g_1^{2}-x^2}\right]
+f_1g_1^{2}\left[\frac{\rad^{\prime}\left(\partial_t f_1+\rad^{\prime}\partial_r f_1\right)+f_1 \rad^{\prime\prime}}{g_1^{2}-x^2}\right]+Gg_1-Fx\right\},
\label{eq:genforcer0}
\end{equation}
\end{widetext}
where all quantities are calculated at $r=\rad(t)$.  We note that, as
required, in the case of a static boundary, for which $\rad(t)=\rad$,
this force reduces to that required to keep an object at rest, which
is given in (\ref{eq:force}), calculated at the boundary.  We may
calculate specific forms for $f_{\rad(t)}$ using either the exterior
solutions given in \cite{NLH1}, or the interior solutions derived here
and given in Table \ref{table:intparams}.


For a spatially-flat background, the force required to keep a test
particle on the boundary of the object is found to be
\begin{equation}
f^{\{k=0\}}_{\rad(t)}=\frac{m}{\rad^{2}(t)\sqrt{1-\frac{2m}{\rad(t)}}}.\label{eq:intforcebound}
\end{equation}
Unlike the force required to keep a test particle at rest in the
object's interior, this force is {\itshape{independent}} of the
acceleration/deceleration $q(t)$ of the background expansion.  It is,
however, still dependent on the rate of expansion/contraction through
the function $\rad(t)$. We point out that the force
(\ref{eq:intforcebound}) has the same the form as that required to
keep a particle at fixed $r$ in the Schwarzschild geometry exterior to
a static, spherical mass $m$ \cite{GGTGA}, but with $r$ replaced by
$\rad(t)$.  Interestingly, we found an analogous modification to
Buchdahl's theorem in equation (\ref{eq:buchdahl}).  These are perhaps
manifestations of Birkhoff's theorem which states that, without
demanding spacetime to be static or stationary, the geometry outside a
general spherically-symmetric matter distibution is the Schwarzschild
geometry \cite{GRbook}.  Since Birkhoff's theorem does not
specifically take an expanding background into account however, a full
investigation into this analogy is left for future research.  For
comparison with results for spatially-curved backgrounds given below,
we note that expanding (\ref{eq:intforcebound}) in small quantities
within the region $m\ll r\ll \rad(t)\ll 1/H_{\rm e}(t)$, and keeping all
terms up to second order, gives
\begin{equation}
f^{\{k=0\}}_{\rad(t)}\approx \frac{m}{\rad^{2}(t)}+\frac{m^2}{\rad^{3}(t)}.\label{eq:frotexpand}
\end{equation}

For a spatially-curved background, although it is difficult to obtain an
analytic expression for the force (\ref{eq:genforcer0}), expanding the
expression in small quantities, and keeping all terms up to second
order, one obtains
\begin{equation}
f^{\{k=\pm1\}}_{\rad(t)}\approx\frac{m}{\rad^{2}(t)}+\frac{3k}{2}\frac{m}{R^2(t)}+\frac{m^2}{\rad^{3}(t)}.\label{eq:intforceboundcurved}
\end{equation}
Compared with the spatially-flat case, this expression contains an
extra contribution proportional to $m/R^2(t)$.  This may be
interpreted as a type of anti-gravitational/gravitational force, for
$k=-1$ and $k=1$ respectively, felt by the test particle if it were
dragged out to the curvature scale of the universe and the central
mass were pointlike.

\section{Discussion and conclusions}\label{sec:conc}

In \cite{NLH1} we derived metrics describing the spacetime around a point
mass embedded in each of a spatially-flat, open and closed expanding universe. 
In each case, we also derived the corresponding
invariant expression for the force required to keep a test particle in
that spacetime at rest, i.e.  at a fixed physical radial coordinate
$r$ relative to the point mass.  In \cite{NLH2}, we used these models to
investigate the effects of universal expansion on astrophysical
objects, such as galaxies and galaxy clusters.

In reality, however, astrophysical objects are of finite spatial
extent. In this paper, we therefore extend our previous work by
considering a finite object, modelled as a spherical `interior' region
of uniform density $\rho_{\rm i}(t)$, embedded in an expanding `exterior'
background, also of uniform density $\rho_{\rm e}(t)$. In each region we
assume a single `phenomenological' fluid that supports pressure,
envisaged to comprise of ordinary gas pressure due to baryonic matter
and an effective pressure from dark matter arising from phase-mixing
and relaxation. In general, the object is dynamic, having a
time-dependent boundary $\rad(t)$, and the expansion rates of the two
regions, expressed in terms of interior and exterior Hubble parameters
$H_{\rm i}(t)$ and $H_{\rm e}(t)$, respectively, can be independent.  For a given
exterior cosmology, one may specify the dynamics of the model by the
choice of the internal Hubble parameter $H_{\rm i}(t)$ (together with the
radius $a_\ast\equiv a(t_\ast)$ and density $\rho_{\ast}\equiv
\rho_{\rm i}(t_\ast)$ of the object at some reference time $t=t_\ast$). In
principle, $H_{\rm i}(t)$ can take any form (and be positive or negative),
since the interplay between pressure and self-gravity in the object
may allow it to expand or contract at any rate. Our model satisfies
the boundary conditions that the fluid velocity, acceleration and
pressure are continuous across the object boundary. In particular, the
continuity of the fluid velocity means that matter does not cross the
boundary in either direction.

By making appropriate choices for the parameters in the interior and
exterior regions, our model could be used to study a wide range of
physical systems, beyond simply an object embedded in the general
fluid of the `expanding universe'.  For example, one could use it to
model a star in a galaxy, a galaxy inside a cluster, or a cluster
inside a supercluster. It could even be be used to model large-scale
inhomogeneties in the universe.  In this context, it may prove useful
in investigating whether local inhomogeneities could provide an
explanation for the observations of the acceleration of the universal
expansion, without invoking dark energy. This might occur if we, as
observers, reside in an `interior' part of the universe that happens
to be less dense, and therefore expanding faster, than the region
exterior to it. By observing a source in the exterior region, one
would then measure an apparent acceleration of the universe's
expansion, but this would be only a local effect.  Although the
magnitude of such effects are likely to be small, it is worth
investigating this quantitatively, and we plan to carry this out in a
future publication.

In the current work, we first perform a Newtonian analysis of our model system, and derive
full analytical solutions for the various physical parameters
describing it, such as the fluid velocities, densities and pressures
in both the interior and exterior regions, and the gravitational
potential in each region. We believe that these solutions have not
appeared previously in the literature.

We then undertake a fully general-relativistic analysis of our model,
employing a tetrad-based procedure for general spherically-symmetric
systems. We first, however, use this approach to derive a generalised
form of the Oppenheimer--Volkov equation, which applies to general
dynamical spherical systems and is of interest in its own right. We
show that in the special case of a static spherical system, our equation
reduces to the standard Oppenheimer--Volkov equation. In the
subsequent general-relativistic analysis of our model, we obtain
analytical solutions for most of the relevant quantities defining the
line-element in the interior and exterior regions, respectively. Some
quantities cannot be found analytically, however, so we give the
corresponding differential equations that must be integrated
numerically to obtain them.

We investigate two interesting special cases of our model: an object
with a static boundary, $H_{\rm i}(t)=0$; and an object whose internal
Hubble parameter matches that of the background, $H_{\rm i}(t)=H_{\rm e}(t)$.  In
the first case, we find that the total energy (mass) $M$ of the object
must remain constant, but its excess mass $m(t)$ (i.e. that in
addition to what would be present due to the background alone) is
still time-dependent.  We focus particularly on calculating the form
of the radial pressure profile in both the interior and exterior
regions.  Our second special case corresponds to an object with
constant excess mass $m$, but time-dependent total energy (mass)
$M(\rad(t),t)$, and so is complementary to our first case.  We
concentrate on deriving forms for the spacetime metric in the two
regions, in the case of a spatially-flat, open and closed background,
respectively. In the exterior region, we recover the solutions
obtained in \cite{NLH1}. In the interior region, we find an analytical
form for the metric in the case of a spatially-flat background. We
show that this metric is, in fact, equivalent to that previously
derived by Nolan \cite{nolan} in isotropic and comoving
coordinates. For the spatially-curved backgrounds, the coefficients in
the interior metric contain a function expressible only as an elliptic
integral. In these cases, we therefore obtain approximate series
solutions for the metric coefficients.  For all three background
geometries, we also obtain expressions for the force required to keep
a test particle at rest (i.e. with fixed physical coordinate $r$) in
the interior of the object, and that required to keep a test particle
on the moving boundary $r=\rad(t)$ of the object.  In the
spatially-flat case, the latter force takes on the same form as that
required to keep a test particle at rest in the Schwarzschild geometry
outside a static mass $m$, but with $r\rightarrow \rad(t)$.  This
modification mirrors a similar modification found to Buchdahl's
theorem in Section \ref{sec:nolan}, which may be a manifestation of
Birkhoff's theorem.  This topic is worthy of further consideration.

It is worth mentioning a recent relevant publication by Zhang \& Yi
\cite{zhang}.  In this work the authors highlight that, for a model
which has both an interior and an exterior region, contrary to popular
belief, the exterior spacetime does have an effect on the interior.
Not including this effect is common, but incorrect, and arises from a
misunderstanding of Birkhoff's theorem.  Our work provides an elegant
demonstration of this fact.  By dealing with the interior and exterior
regions simultaneously in our analyses, and ensuring that physical
quantities match at the boundary of the object, we have seen that that
exterior parameters do indeed appear in our interior expressions. For
example, the rate of expansion of the boundary of an object
$\rad^{\prime}(t)$ is in general determined by both $H_{\rm i}(t)$ and
$H_{\rm e}(t)$, as is evident from equation (\ref{eq:r1prime}).

Finally, we must mention that our model has some obvious
oversimplifications, most notably the assumption of uniform density in
both the interior and exterior regions. In principle, our approach
could be extended to accommodate a general radial density profile,
which would also obviate the need to consider two separate regions.
In particular, it would be of interest to investigate the evolution of
an object described by a more realistic density profile for, say, a
cluster of galaxies. Such considerations are left for future research.

\begin{acknowledgments}
RN was supported by a Research Studentship from the Science and
Technology Facilities Council (STFC).
\end{acknowledgments}

\appendix
\section{Generalised Oppenheimer--Volkov equation for
a point mass in an expanding universe}\label{sec:OVpointmass}

We illustrate here the use of our generalised Oppenheimer--Volkov
equation (\ref{eq:OVgeneralised}) by applying it to the special case
of a point mass in a homogeneous and isotropic expanding universe,
as studied in \cite{NLH1}.

For a point mass, one has $M=\frac{4}{3}\pi\rho_{\rm e}(t)r^3+m$ and we
found in \cite{NLH1} that $v=rH_{\rm e}(t)$.  Using the exterior
cosmological equations (\ref{eq:cosmo1})--(\ref{eq:cosmo3}), it can be
shown that the generalised Oppenheimer--Volkov equation
(\ref{eq:OVgeneralised}) for this system is
\begin{widetext}
\begin{equation}
\partial_r p_{\rm e}
=-\frac{\rho_{\rm e}(t)+p_{\rm e}}{r\left(1-\frac{2m}{r}-\frac{kr^2}{R^2(t)}\right)}
\left[\frac{m}{r}+\left(\frac{p_{\rm e}-p_\infty(t)}{\rho_{\rm e}+p_\infty(t)}\right)
\frac{kr^2}{R^2(t)}\right].
\end{equation}
\end{widetext}
where the purely time-dependent pressure $p_\infty(t)$ corresponds to
the external cosmological fluid. This result, valid for general $k$,
was not previously published.  Although it cannot be integrated
analytically for general $k$, in the spatially-flat case ($k=0$) it
yields the analytical result
\begin{equation}
p_{\rm e}=\rho_{\rm e}(t)\left[\frac{1+w_\infty}{\sqrt{1-\frac{2m}{r}}}-1\right].
\end{equation}
where $w_\infty \equiv p_\infty(t)/\rho_{\rm e}(t)$ is the equation-of-state
parameter for the cosmological fluid, which is time-independent for
standard fluids such as dust $(w_\infty=0)$ and radiation
$(w_\infty=\tfrac{1}{3})$.  Indeed, the above expression agrees with
the result obtained in equation (22) in \cite{NLH1}.

\section{Relativistic junction conditions at the object boundary}

The line-element describing our spherically-symmetric system has the
form given in (\ref{eq:generalmetric}). The 3-dimensional timelike hypersurface traced
out by the (in general) moving spherical boundary of the object is
defined by $\Sigma(t,r) \equiv r-a(t) = 0$. The unit (spacelike)
normal to $\Sigma$, pointing from the interior to the exterior region,
is easily found to have the covariant components
\begin{equation}
[\hat{n}_\alpha] = \frac{[-a'(t), 1, 0, 0]}
{|f_1^2 {a'(t)}^2-2f_1 g_1 a'(t) + g_2^2-g_1^2|^{1/2}},
\label{eq:nhatgen}
\end{equation}
from which one may readily verify that $\hat{n}_\alpha \hat{n}^\alpha
= -1$.

The first Israel junction condition requires continuity of the induced
metric on $\Sigma$, which is given by
\begin{equation}
h_{\mu\nu} = g_{\mu\nu} +\hat{n}_\mu \hat{n}_\nu,
\label{eq:inducedmetric}
\end{equation}
where $g_{\mu\nu}$ can be read off from
(\ref{eq:generalmetric}). In addition, $h_{\mu\nu}$ acts as a projection tensor onto
the hypersurface $\Sigma$. One sees immediately that, for the induced
metric to be continuous at the object boundary, one requires the tetrad
components $f_1$, $g_1$ and $g_2$ all to be continuous there, which
thus constitute our first set of boundary conditions.

In the analysis of our model system presented in
Section~\ref{sec:generalresults}, one finds that continuity of $f_1$
and $g_2$ at the boundary implies that the object radius changes according
to $L_ta(t)=a(t)H_{\rm i}(t)$, which is equivalent to $a'(t) =
[g_2/f_1]_{\rm b}$, where the right-hand side is evaluated at the
boundary. Thus, on the hypersurface $\Sigma$, one may write $a'(t) =
g_2/f_1$ so that the expression (\ref{eq:nhatgen}) for the unit normal
simplifies to
\begin{equation}
[\hat{n}_\alpha] = \frac{1}{g_1}[-g_2/f_1, 1, 0, 0],
\label{eq:nhatdown}
\end{equation}
where we have also used the fact that $g_1$ may always be taken as
positive. Moreover, the contravariant components of the unit normal
have the simple form
\begin{equation}
[\hat{n}^\alpha] = [0, -g_1, 0, 0].
\label{eq:nhatup}
\end{equation}

The second Israel junction condition requires continuity of the
extrinsic curvature of $\Sigma$, which is given (up to a sign
ambiguity) by
\begin{align}
K_{\alpha\beta} &= {h_\alpha}^\mu {h_\beta}^\nu \nabla_\mu \hat{n}_\nu \nonumber \\
&= {h_\alpha}^\mu \nabla_\mu \hat{n}_\beta \nonumber \\
& = \nabla_\alpha \hat{n}_\beta
+ \hat{n}_\alpha \hat{n}^\mu \nabla_\mu \hat{n}_\beta,
\end{align}
where we have used (\ref{eq:inducedmetric}) and the fact that $n^\mu
n_\mu =-1$. Using the expressions (\ref{eq:nhatdown}) and
(\ref{eq:nhatup}) for the components of the unit normal, the extrinsic
curvature thus has the simple form
\begin{widetext}
\begin{equation}
K_{\alpha\beta} = \partial_\alpha\hat{n}_\beta-g_1\hat{n}_\alpha
\partial_r\hat{n}_\beta+
\frac{1}{g_1}\left(\frac{g_2}{f_1}{\Gamma^0}_{\alpha\beta}-{\Gamma^1}_{\alpha\beta}\right)-\hat{n}_\alpha\left(\frac{g_2}{f_1}{\Gamma^0}_{1\beta}-{\Gamma^1}_{1\beta}\right).
\end{equation}
\end{widetext}
Substituting for the components of the unit normal using the
expressions (\ref{eq:nhatdown}) and (\ref{eq:nhatup}), and calculating
the necessary connection coefficients of the line-element
(\ref{eq:generalmetric}), after a lengthy, but straightforward,
calculation, one finds that the only non-zero components of the
extrinsic curvature are
\begin{align}
K_{00} & = \frac{g_2^4-g_1^4+g_1^2g_2^2}{f_1^3g_1g_2^2} \partial_r f_1, \nonumber \\
K_{22} & = g_1r,\nonumber \\
K_{33} & = g_1r\sin^2\theta.
\end{align}
Since the first Israel junction condition requires $f_1$, $g_1$ and
$g_2$ all to be continuous at the object boundary, then the second
junction condition requires only that, in addition, $\partial_r f_1$
is continuous there.

\bibliography{references}

\end{document}